\documentclass{vgtc}                          

\graphicspath{{figures/}{pictures/}{images/}{./}} 

\usepackage{times}                     

\usepackage{tabu}                      
\usepackage{booktabs}                  
\usepackage{lipsum}                    
\usepackage{mwe}                       
\usepackage{cancel}
\usepackage{comment}

\usepackage{algorithm}

\usepackage{mathptmx}                  

\usepackage{booktabs}
\usepackage{siunitx}  

\usepackage{graphicx}
\usepackage{subcaption}

\usepackage{graphicx} 
\usepackage{amsmath} 
\usepackage{amsthm}  
\usepackage{amssymb}  
\usepackage{algorithm} 
\usepackage{algpseudocode}

\usepackage{xcolor} 
\usepackage{wrapfig} 

\newtheorem{theorem}{Theorem}  

\newcommand{\ve}  {{\varepsilon}}
\newcommand{\tfp}  {{T_1^{f,\pi_1}}}
\newcommand{\tgp}   {{T_2^{g,\pi_2}}}
\newcommand{\StateTwoP}		{\texttt{S2\_particle}\xspace} 
\newcommand{\StateOneP}  {\texttt{S1\_particle}\xspace} 
\newcommand{\StateOneH}  {\texttt{S1\_hole}\xspace} 
\newcommand{\StateTwoH}  {\texttt{S2\_hole}\xspace} 
\newcommand{\MethodOne}  {{Method 1}\xspace} 
\newcommand{\MethodTwo}  {{Method 2}\xspace}

\newcommand{\revision}[1]{\textcolor{black}{#1}}

\onlineid{1011}
\vgtccategory{Research}
\vgtcinsertpkg

\title{Efficient Heuristic Algorithms for Interleaving Distance between Merge Trees}

\author{Elena Farahbakhsh Touli\thanks{e-mail: elena.farahbakhsh.touli@liu.se}
\and Talha Bin Masood\thanks{e-mail: talha.bin.masood@liu.se}}
\affiliation{\scriptsize Scientific Visualization Group, Department of Science and Technology (ITN), Link\"oping University, Norrk\"oping, Sweden}

\abstract{
Merge trees are fundamental structures in topological data analysis. Interleaving distance is a widely accepted metric for comparing merge trees, with applications in visualization and scientific computing.  While a greedy algorithm exists for finding the interleaving distance between labeled merge trees with overlapping labels, computing the interleaving distance between unlabeled trees or labeled trees with disjoint labels remains a significant challenge.

In this work, we introduce a novel heuristic algorithm for approximating the interleaving distance between labeled merge trees with partial agreement and disagreement. Our method strategically assigns labels primarily to the leaves of the trees to infer structural correspondence. We also introduce an enhanced version of a previous algorithm that offers improved performance. Both algorithms run in polynomial time and provide practical, efficient alternatives for comparing merge trees, particularly in cases involving unlabeled or structurally diverse data. This work contributes a new direction for merge tree analysis and offers promising tools for real-world applications. We demonstrate this application on the simulation of time-varying electron density.
} 


\begin{document}


\firstsection{Introduction}

\maketitle
Merge trees are topological structures that describe how the connected components of sublevel sets of a real-valued function evolve over a topological space~\cite{carr2003computing}. They are widely used in topological data analysis because they provide a compact and interpretable summary of the shape of the data~\cite{heine2016survey}. Merge trees play a critical role in a range of applications, including visualization \cite{MSDiSFbCC,ASAoLMTfUV}, chemistry \cite{BToDL, EEDEUMTM}, physics~\cite{Thygesen2023}, 
data simplification, shape comparison~\cite{biasotti2008describing}, and scientific computing, where capturing the hierarchical structure of features in a scalar field is essential \cite{SiSFT}.


To compare merge trees quantitatively, \emph{interleaving distance} has been introduced as a metric. The original definition of interleaving distance was proposed by Chazal et al. in \cite{PoPMaTD}, and then was extended by Morozov et al.~\cite{IDbMT} for finding the distance between merge trees. They use two $\ve$-compatible maps to measure similarity between merge trees. Later, Agarwal et al.~\cite{CtGHDfMT} explored the relationship between the interleaving distance and the Gromov–Hausdorff distance for metric trees. Subsequently, Touli et al.~\cite{FAfCGHaIDbT} simplified the definition by using a single map, called an $\ve$-good map, and introduced the first fixed-parameter tractable (FPT) algorithm for computing the interleaving distance between merge trees. Their work also provided an FPT approximation algorithm for the Gromov–Hausdorff distance between metric trees.

The notion of interleaving distance was further extended to \emph{labeled merge trees} \cite{IIDfMT,TeiCMfPTaaID, ASAoLMTfUV, yan2022geometry}, where a label map is defined on the vertices of the tree, such that each leaf is assigned at least one label. The authors demonstrated that the interleaving distance could still be computed in this setting and proposed a greedy algorithm for trees that share at least one label. In their method, for any leaf in the tree with the most unmatched labels, a corresponding leaf is selected from another tree. However, based on the definition of interleaving distance, there are some leaves that are mapped to an inner node. 
In addition, when trees are unlabeled or have completely disjoint labels, their approach requires additional information about the embedding of the trees in $\mathbb{R}^d$,
which introduces additional complexity. Thus, computing the interleaving distance between unlabeled merge trees -- or labeled trees with no common labels -- remains a challenging and largely unresolved problem.

Our work addresses these gaps. We propose a heuristic, polynomial-time algorithm for computing the interleaving distance between unlabeled merge trees.
This approach differs fundamentally from previous methods and provides a practical solution for a broader range of tree comparison problems. In addition, we make a slight modification to the previous method and propose a heuristic algorithm that achieves better performance.
\\
\\
{\textbf{Contributions.}}
In this paper, we introduce two new methods for computing the interleaving distance between merge trees using a heuristic labeling strategy. A key advantage of our approaches is that they eliminate the need for information about embedding trees in any geometric space. Instead, we focus primarily on labeling the leaves of the trees, enabling efficient computation of the interleaving distance while preserving meaningful structural information.

In Section 2, we provide the necessary background and discuss the existing greedy method. In Section 3, we illustrate our two proposed methods using examples. Section 4 presents the experimental results, where we evaluate our methods. The final section offers discussion and conclusions.
\\
\\
{\textbf{Related Work.}}
The problem of measuring distances between trees has received considerable attention in recent research. 
One of the metrics that can be used to compute the distance between trees is the \textit{Hausdorff distance}. However, the Hausdorff distance does not effectively capture structural similarity between trees. In fact, it can assign a small distance even to trees with significantly different topologies \cite{GMMFaSA}. 

Other distances that have been used in the distance between rooted trees are \textit{edit distance} and \textit{alignment distance}~\cite{ASoTEDaRP}. It has been shown in~\cite{AoTAAtTE, SMSRCULT} that computing both the tree edit distance and the tree alignment distance for arbitrary trees is \emph{MAX SNP-hard}. However, there exist polynomial-time algorithms for specific cases: the tree edit distance can be computed efficiently between two labeled, ordered trees of bounded depth, and the tree alignment distance can be computed in polynomial time between any two labeled, ordered trees with bounded vertex degrees~\cite{AoTAAtTE}. In practice, it is often difficult to restrict the depth of trees, but assuming bounded vertex degrees is quite common. Although a polynomial-time algorithm exists for computing the tree alignment distance in this setting, we do not focus on it here, as the tree alignment distance heavily depends on the specific parent-child relationships, which limits its applicability in our context~\cite{GMMFaSA}. Both tree edit distance-based methods~\cite{sridharamurthy2018edit,pont2021wasserstein,wetzels2022branch} and tree alignment-based methods~\cite{lohfink2020fuzzy} have been proposed for merge trees. We refer the reader to a comprehensive survey by Yal et al.~\cite{yan2021scalar} on comparison measures for topological structures including merge trees.  

Another distance defined between pairs of metric spaces is the \textit{Gromov–Hausdorff distance}. However, it has been proven that there is no polynomial-time algorithm capable of approximating the Gromov–Hausdorff distance between two metric trees within a factor better than 3, unless \(\text{P} = \text{NP}\) \cite{CtGHDfMT, OtUoGHdfSC, CAotGHDaiAiNSM}.


\section{Preliminaries}
A \emph{graph} $G = (V,E)$ is a mathematical structure used to model pairwise relationships between objects. Here, $V$ represents the set of vertices, and $E$ represents the set of edges. A connected graph is a graph in which there is a path between any pair of vertices in the graph. A connected graph that contains no cycles is called a tree. A rooted tree is a tree in which one vertex, say $v$, is designated as the root. In a rooted tree, there is a parental relationship between the vertices connected by edges, where the vertex closer to the root is considered the parent of the other vertex. In rooted trees, each vertex has an \emph{indegree} and an \emph{outdegree}. The indegree is always $1$ for vertices other than the root (the indegree of the root is $0$), while the outdegree indicates the children of each vertex. Vertices that are not the root and have an outdegree greater than $0$ are called internal vertices, while non-root vertices with an outdegree of $0$ are referred to as leaves \cite{bondy}.

A \emph{metric space} $\mathcal{X}$  is a pair of $(X,d_\mathcal{X})$ where $X$ is a set and $d_{\mathcal{X}}:(x,y)\rightarrow \mathbb{R}_{\geq 0}$ is a metric satisfying the following properties for every $x,y,z \in X$: 
\begin{align*}
&\text{(1)} \quad\text{Non-negativity}\quad d_{\mathcal{X}}(x, y)\geq 0 \\
&\text{(2)} \quad\text{Identity} \quad d_{\mathcal{X}}(x, y)=0, ~\text{if and only if} ~ x = y \\
&\text{(3)} \quad\text{Symmetry} \quad d_{\mathcal{X}}(x, y) = d_{\mathcal{X}}(y, x) \\
&\text{(4)} \quad\text{Triangle inequality} \quad d_{\mathcal{X}}(x, y) \leq d_{\mathcal{X}}(x, z) + d_{\mathcal{X}}(z, y).
\end{align*}

A metric space $\mathcal{X}$  is called a finite metric tree \cite{FAfCGHaIDbT} if it satisfies two conditions: first, $X$ is a length metric space; and second, $X$ is homeomorphic to the geometric realization \( |X| \) of a finite tree $X = (V, E)$. Therefore, we show the metric tree $\mathcal{T}$ with $(|T|, d_\mathcal{T})$, where $|T|$ is the underlying space of $T = (V, E)$.

A \emph{merge tree} 
is defined as a rooted tree $T$ and a function $f$ which is defined on the geometric realization of the tree $|T|$. The function $f$ is decreasing from the root to the leaves. Intuitively, a merge tree can be constructed from any given weighted tree (with positive edge weights) by selecting a vertex $u$ and \emph{hanging} the tree from $u$ near the real line. The vertex $u$ is assigned a function value of $0$, and the function values of the other vertices are determined by their positions along the real line, based on the negative accumulated edge weights from $u$. Additionally, an edge is added from $u$ to infinity. In this paper, we focus exclusively on finite merge trees, that is, trees consisting of a finite number of vertices.

\subsection{Unlabeled Merge Tree.}
For a given space $\mathcal{X}$ with a function $f:\mathcal{X} \rightarrow \mathbb{R}$, we define $x\sim y$ if:  
(1) $f(x) = f(y) = a$, and  
(2) $x$ and $y$ are in the same connected component of $f^{-1}(-\infty,a]$.  
Then the \emph{quotient space} $\mathcal{X} \setminus \sim$ defines a merge tree \cite{FAfCGHaIDbT}. In other words, a merge tree $T^f$ is a pair $(T,f)$ consisting of a rooted tree $T$ with an underlying space $|T|$ and a function $f:|T|\rightarrow \mathbb{R} \cup \{\infty\}$, where the function is decreasing from the root to the leaves. To illustrate further, consider a weighted tree $T = (V, E, W_{> 0})$ which $V$ is a set of vertices and $E$ a set of edges and $W: E\rightarrow \mathbb{R}_{> 0}$ assigns a positive weight (or length) to each edge. The underlying space of $T$, denoted by $|T|$ is the topological space constructed by representing each edge as a closed interval — typically $[0, W(e)]$ (if the tree $T$ is unweighted, each edge is identified with the unit interval [0,1]). And gluing these intervals together at the vertices according to the connectivity defined by $E$. Moreover, $f(u) = \infty$ if and only if $u$ is the root of the tree.

To compare merge trees both structurally and functionally, the concept of \emph{interleaving distance} is often employed. Specifically, we say that two continuous maps, $\alpha:|T_1^f| \rightarrow |T_2^g|$ and $\beta:|T_2^g| \rightarrow |T_1^f|$, $\varepsilon$-interleave if and only if, for every pair of points $u \in |T_1^f|$ and $w \in |T_2^g|$, the following four conditions are satisfied~\cite{IDbMT}:
\begin{align*}
&\text{(1)} \quad g(\alpha(u)) = f(u) + \varepsilon  \quad\quad\quad
&\text{(2)}\quad \beta(\alpha(u)) = u^{2\varepsilon}\\
&\text{(3)} \quad f(\beta(w)) = g(w) + \varepsilon \quad\quad\quad &~~\text{(4)}\quad \alpha(\beta(w)) = w^{2\varepsilon}
\end{align*}
Here, $u^{2\varepsilon}$ denotes the ancestor of $u$ that is located $2\varepsilon$ above it in the tree, meaning $f(u^{2\varepsilon}) - f(u) = 2\varepsilon$. A similar definition applies for $w^{2\varepsilon}$.

\begin{figure}[!htb]
\begin{center}
\centering
\includegraphics[width=8cm, trim=30 300 20 20, clip]{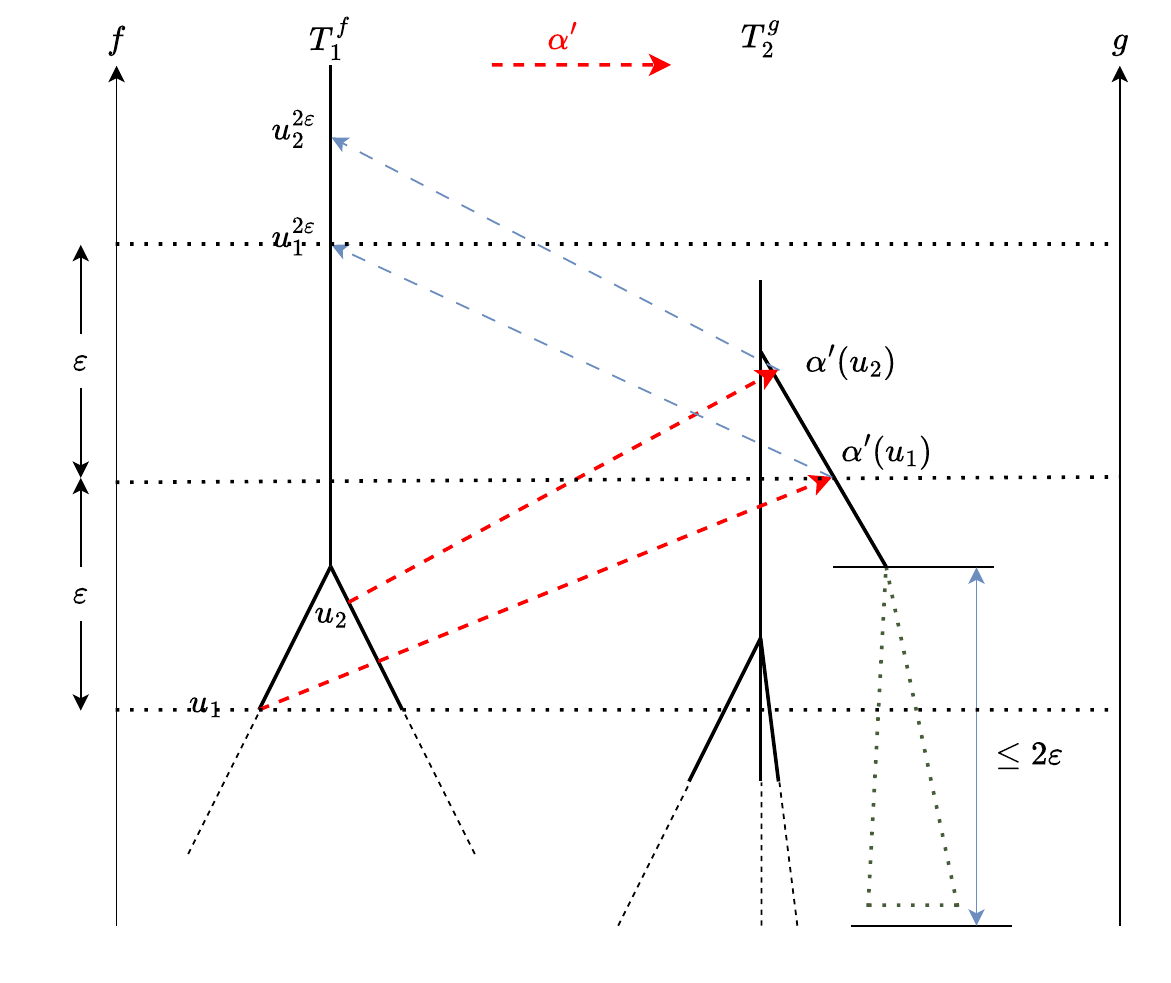}
\caption{The two merge trees $T_1^f$ and $T_2^g$ are compared using an $\varepsilon$-good map $\alpha' : |T_1^f| \rightarrow |T_2^g|$. The green subtrees in $T_2^g$ represent regions in $T_2^g \setminus \text{Img}(\alpha')$, where each region consists of points whose distance to their nearest ancestor within the image of $\alpha'$ is at most $2\ve$. For example, in this figure, $\alpha'(u_2)$ is an ancestor of $\alpha'(u_1)$, and consequently, $u_2^{2\ve}$ is an ancestor of $u_1^{2\ve}$, illustrating the second condition.}
\label{fig:einterleaving}
\end{center}
\end{figure}

In \cite{FAfCGHaIDbT}, it was shown that the interleaving distance between merge trees can be computed using a single map instead of two. This map is referred to in this work as an $\ve$-good map. This approach enables the development of an FPT algorithm for computing the interleaving distance between certain merge trees. Given two merge trees $T_1^f$ and $T_2^g$, an $\ve$-good map $\alpha':|T_1^f| \rightarrow |T_2^g|$ is a continuous map that satisfies the following three conditions: 
\begin{align*}
&\text{(1)} \quad g(\alpha(u)) = f(u) + \varepsilon \\
&\text{(2)} \quad {\text{if }} ~ \alpha(u_1) \preceq \alpha(u_2), \quad u_1^{2\varepsilon} \preceq u_2^{2\varepsilon} \\
&\text{(3)} \quad \text{if } ~ w \in T_2^g\setminus \text{Img}(\alpha), \quad g(w^f) \leq g(w) + 2\ve
\end{align*}
where $w^f$ denotes the nearest point to $w$ that lies in $Img(\alpha)$ and $u_1 \preceq u_2$ indicates that $u_2$ is either equal to $u_1$ or is an ancestor of $u_1$ (see \cref{fig:einterleaving}).


\begin{theorem}\cite{FAfCGHaIDbT}\label{theo:Touli}
For given merge trees $T_1^f$ and $T_2^g$, there exists an $\ve$-good map $\alpha':|T_1^f| \rightarrow |T_2^g|$, if and only if there exist a pair of $\ve$-interleave maps $\alpha:|T_1^f| \rightarrow |T_2^g|$ and $\beta:|T_2^g| \rightarrow |T_1^f|$. 
\end{theorem} 
This theorem states that the interleaving distance can be computed using a single map, known as an $\ve$-good map. 

\subsection{Labeled Merge Trees.} \label{sec:LabeledMergeTree}
The concept of a labeled merge tree was introduced in \cite{TeiCMfPTaaID, IIDfMT, ASAoLMTfUV}.
For a fixed $n$, an $n-$labeled merge tree, denoted as $T^{f,{\pi}}$, is a merge tree equipped with a label map ${\pi}: [n]\rightarrow V(T)$, where $[n] = [1,2,3,..., n]$ and each leaf has at least one label. 
The labeling map ${\pi}$ is not necessarily injective, meaning that a single vertex may have multiple labels. For a labeled merge tree the \emph{induced matrix} $M^{T^{f, {\pi}}}$ is an $n \times n$ symmetric matrix defined as $M^{T^{f, {\pi}}}(i, j) = f(LCA(v_i, v_j))$, where $LCA(v_i, v_j)$ denotes the lowest common ancestor of the vertices $v_i$ and $v_j$.

For two merge trees, $T_1^{f}$ and $T_2^g$, and a given $\ve$, $T_1^{f}$ and $T_2^{g}$ are $\ve$-label if there exists two label maps ${\pi_1}: [n]\rightarrow V(T_1)$ and ${\pi_2}: [n]\rightarrow V(T_2)$ such that $\|M^{T_1^{f, {\pi_1}}}-  M^{{T_2}^{g, {\pi_2}}}\|_{\infty} = \ve$. The label interleaving distance $d_{ID}^L (T_1^f$, $T_2^{g})$
is defined as the infimum value of $\ve$ for which the two labeled merge trees $T_1^{f}$ and $T_2^{g}$ are $\ve$-label.

\begin{theorem}\cite{IIDfMT}
    For a pair of merge trees $T_1^f$ and $T_2^g$, $d_{ID}(T_1^f, T_2^g) = d_{ID}^L(T_1^f, T_2^g)$.
\end{theorem}

Finding the interleaving distance between merge trees is NP-hard to compute \cite{CtGHDfMT}, which is why we are exploring heuristic or greedy algorithms to approximate it.
Yan et al.~\cite{ASAoLMTfUV} introduced a greedy method for computing the interleaving distance between labeled merge trees. Their approach categorizes labeled merge trees into three cases: full agreement, partial agreement, and disagreement. 

\begin{itemize}
    \item \emph{Full agreement} refers to the case where both trees share identical labels.
    \item \emph{Partial agreement} occurs when the trees have different labels but share at least one label in common.
    \item \emph{Disagreement} arises when there are no shared labels between the trees.
\end{itemize}
\revision{Unlabeled merge trees can be viewed as a special case of labeled merge trees, where each leaf is assigned a distinct label -- corresponding to the disagreement case, with the two trees sharing no common labels.}

\begin{figure}[!htb]
\begin{center}
\centering
\includegraphics[width=8cm, trim=20 380 250 20, clip]
{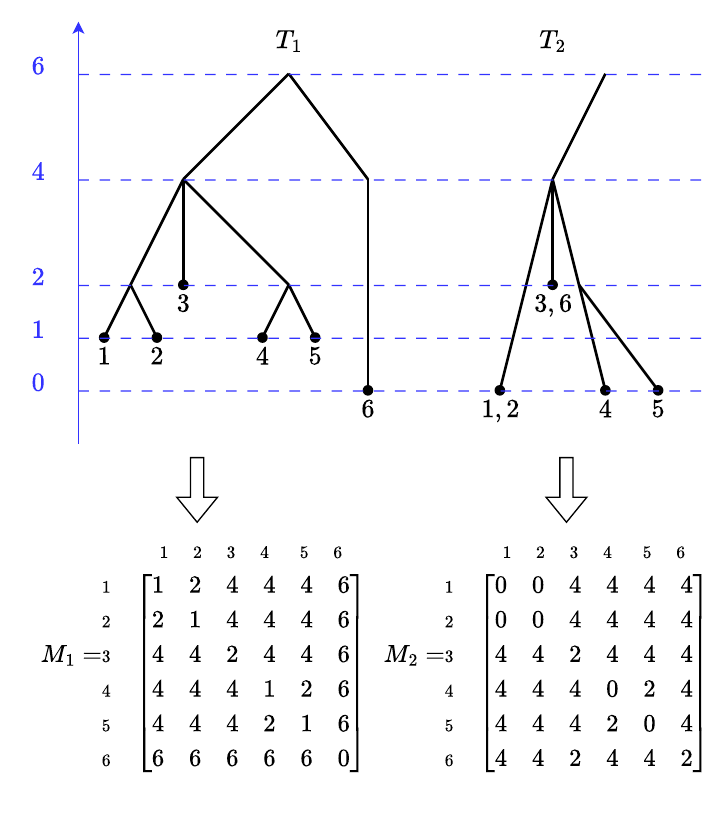}
\caption{The case of full agreement, where both labeled merge trees share the same labels and have corresponding induced matrices.}
\label{fig:fullagreement}
\end{center}
\end{figure}

In the case of full agreement (\cref{fig:fullagreement}), they construct \emph{induced matrices} for both trees (denoted $M^{T_1^{f,{\pi_1}}}$ and $M^{{T_2}^{g,{\pi_2}}}$), and compute the interleaving distance as:
\[
\|M^{T_1^{f, {\pi_1}}} - M^{{T_2}^{g, {\pi_2}}}\|_{\infty}.
\]
In this paper to ease notation, we denote \( M_1 := M^{T_1^{f, \pi_1}} \) and \( M_2 := M^{T_2^{g, \pi_2}} \).


\begin{figure}[!htb]
\begin{center}
\centering
\includegraphics[width=6cm, trim=20 400 20 20, clip]{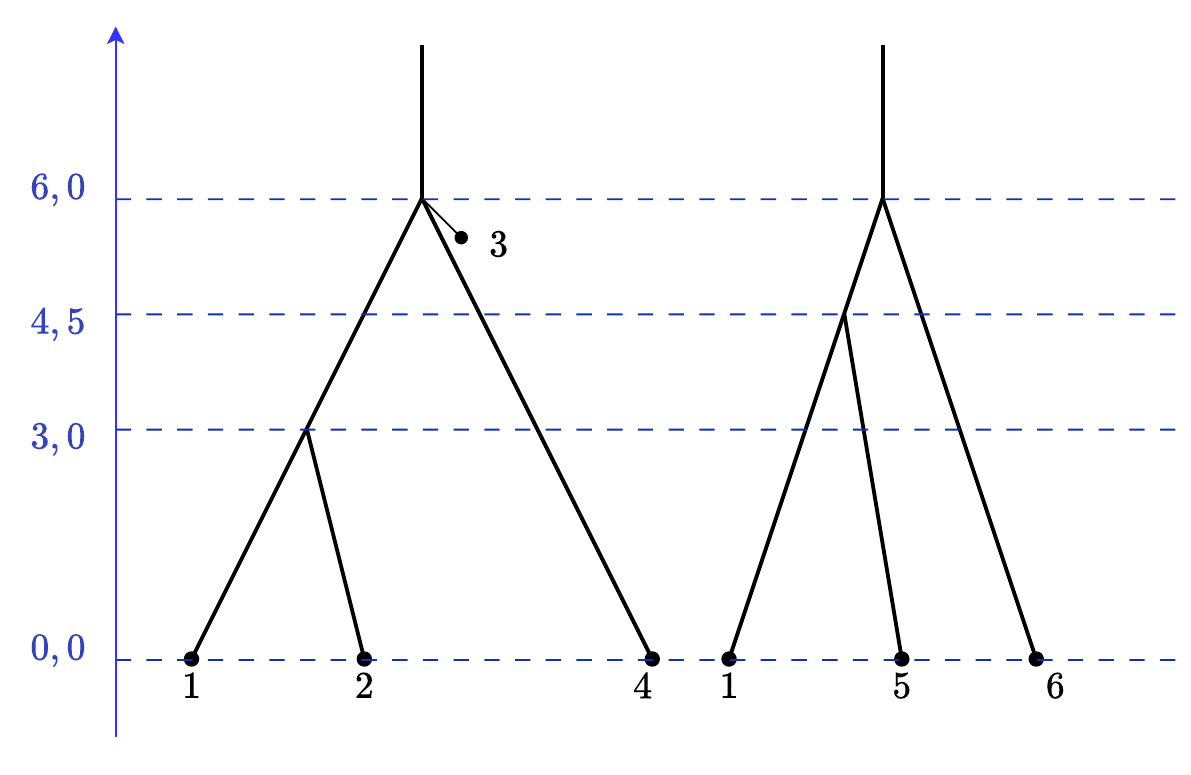}
\caption{Two labeled merge trees with one common known label. Here, vertex $3$ should not be mapped to any leaf in the other tree.}
\label{fig:leaves}
\end{center}
\end{figure}

For partial agreement (\cref{fig:leaves}), the vertices are divided into \emph{known-labeled vertices} (those with matching labels in both trees) and \emph{unknown-labeled vertices} (those without matches). They consider the pivot tree to be the one with the largest number of unknown-labeled vertices. They then compute the distances from each unknown-labeled vertex to all known-labeled vertices, creating a distance matrix for each. Using these matrices, they construct a complete, weighted bipartite graph where the parts correspond to the unknown-labeled vertices from each tree. Each edge $e_{i,j}$ between vertex $v_i$ from tree $\tfp$ and vertex $v_j$ from tree $\tgp$ has a weight defined by:
\[
\|D_1(i) - D_2(j)\|_2
\]
where $D_1(i)$ is the $i-$th row of $D_1$ which is the distance vector from $v_i$ to the ordered known-labeled vertices in tree $\tfp$, and $D_2(j)$ is defined similarly for tree $\tgp$. Next, they apply the {\emph{minimum-weight maximum matching}} algorithm to determine the best correspondence between the main components of the two trees. A \emph{matching} is a subset of a graph's edges such that no two edges share a common vertex. A \emph{minimum-weight maximum matching} is a matching that contains the largest possible number of edges while minimizing the total edge weight \cite{IC}. However, when the two parts of the bipartite graph have unequal numbers of vertices, some vertices in the pivot tree remain unmatched. To handle this, a greedy method is employed to assign labels to the matched vertices. 
To apply the greedy method, for each vertex $v_i$ in the pivot tree, they compute a vector $d_i$, which represents the distances from $v_i$ to the newly known-labeled vertices. These known-labeled vertices consist of the original known-labeled vertices together with the previously unknown-labeled vertices that have been matched through the minimum-weight maximum matching process. In the smaller tree, a distance matrix $D_s$ is computed for its vertices. The vector $d_i$ has the same length as the number of columns in $D_s$. The vertex $v_i$ is then assigned the label of the corresponding vertex $v_{j'}$, where $j'$ is the index that minimizes $\|D_s(j) - d_i\|_2$, and $D_s(j)$ denotes the $j$-th row of the matrix $D_s$.


\begin{figure}[!htb]
\begin{center}
\centering
\includegraphics[width=6cm, trim=20 390 20 20, clip]{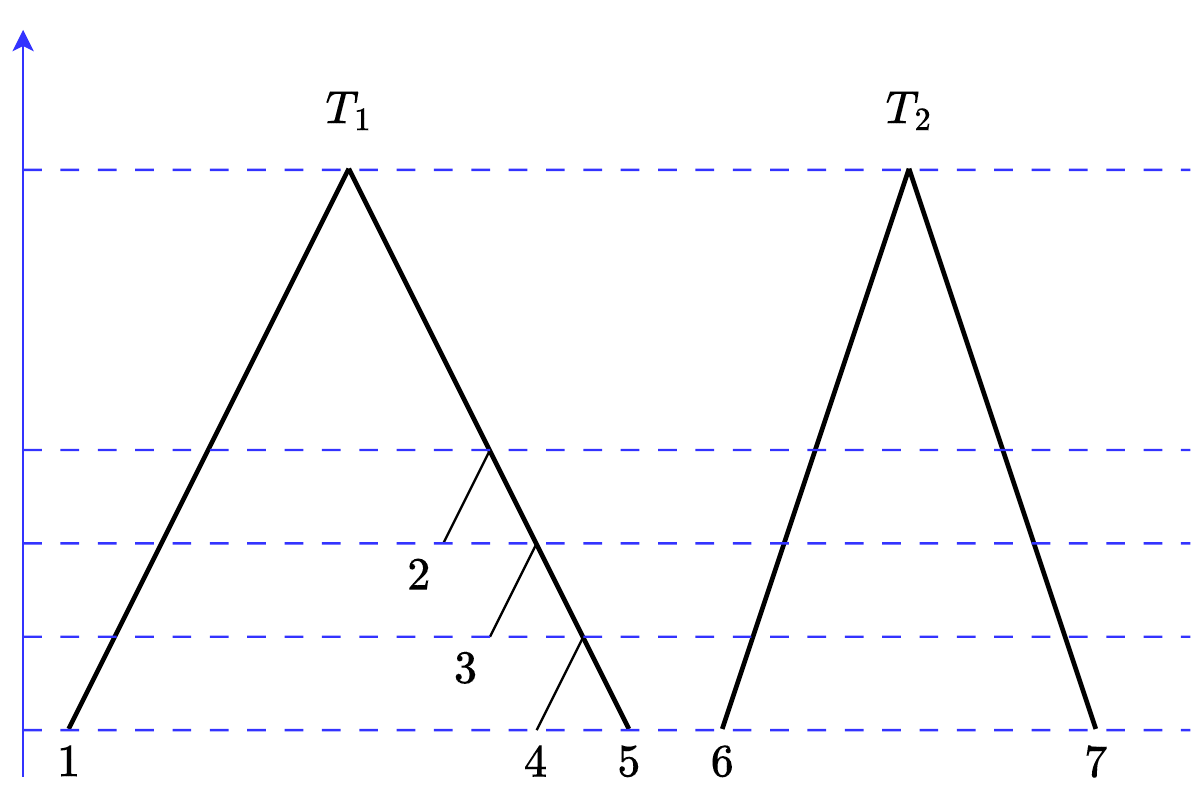}
\caption{An example of disagreement, in which the two labeled merge trees have distinct labels.}
\label{fig:disagreement}
\end{center}
\end{figure}

For the disagreement case (see \cref{fig:disagreement}), the same method is applied, with the Euclidean distance computed from the positions of the vertices in $\mathbb{R}^d$. However, no approach is provided for computing the distance between the trees in the absence of positional information in an external space.

The main difficulty in the case of disagreement is the absence of guiding vertices—such as the known-labeled ones used in the partial agreement case—that are required to construct the distance matrices $D_1$ and $D_2$. This lack of reference points significantly increases the complexity of the computation. 

Additionally, a limitation of the partial agreement scenario is the assumption that leaves must be matched to other leaves. However, as illustrated in \cref{fig:leaves}, trees may contain many short edges that are more appropriately matched to interior points along an edge rather than to specific vertices.


\section{Methods}
This paper presents two heuristic approaches to labeling given merge trees in a way that approximates the interleaving distance, focusing primarily on leaf labeling.

\subsection{Efficient Leaf Matching (\MethodOne)}\label{sec:FasterAlgorithm}
\revision{
The work by Touli et al.~\cite{FAfCGHaIDbT} provides the key intuition for this method. They proved that the interleaving distance between two merge trees can be computed using a single $\varepsilon$-good map instead of two (see \cref{theo:Touli} for further illustration).  
This map is defined on all points of the first tree but is not necessarily surjective, meaning that certain regions of the second tree lie outside the image of the $\varepsilon$-good map.  
To address this, we take the smaller tree as the domain of the map and the larger tree (with more leaves) as its codomain.  
We then identify and effectively trim the portions of the larger tree that fall outside the $\varepsilon$-good map.}

We begin by selecting the tree with the fewest leaves as the \emph{reference tree} (denoted as $T_1$). We then trim the other tree to ensure that both have the same number of leaves. The objective of this trimming step is to remove parts of the tree that are shallow (i.e., have small depths) and lie outside the image of an $\ve$-good map. The depth of a vertex $v$ in the merge tree $T_1^f$ is $f(v)-f(v^d)$ where $v^d$ is the descendant of $v$ with the minimum function value of $f$.

\begin{figure}[!htb] %
\centering
\includegraphics[width=7cm, trim=20 270 330 20, clip]{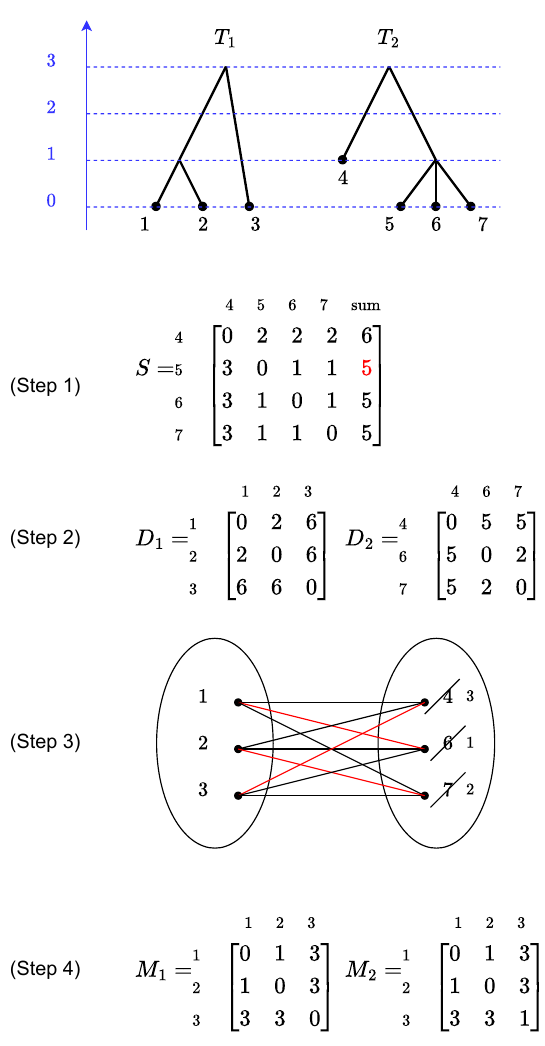}
\caption{Two labeled merge trees, $\tfp$ and $\tgp$, are shown, where the trees have disjoint labels (representing a case of disagreement), and $\tfp$ serves as the reference tree. In Step~1, we construct the matrix $S$ to identify and trim \revision{some leaves}; in this case, the leaf with label~5 is trimmed (three rows have the same sum, and label~5 is selected at random). In Step~2, we compute the distance matrices for each tree. In Step~3, we build the bipartite graph and apply the minimum-weight maximum matching algorithm, resulting in the labels~4, 6, and~7 being switched to~3, 1, and~2, respectively. Finally, in Step~4, we construct the induced matrices.}
\label{fig:firstmethod}
\end{figure}

Next, we apply a minimum-weight maximum matching algorithm to find the best correspondence between the main components of the two trees. The Hungarian algorithm solves the minimum-weight maximum matching problem in 
$O(n^3)$ time, for a given bipartite graph $K_{n,n}$ \cite{IC}. 

Following the approach of \cite{ASAoLMTfUV}, we consider two types of label agreement: partial agreement and disagreement. We exclude the case of full agreement, as it can be handled using the same methodology as in previous works.

In the case of \emph{disagreement}, we first trim the tree. To do this, we construct a nonsymmetric matrix $S$, where each element is defined as $S(i, j) = f(\text{LCA}(v_i, v_j)) - f(v_i)$. We then add a column to this matrix representing the row-wise summation. The $n_2 - n_1$ vertices (where $n_1$ is the number of leaves in tree $\tfp$ and similarly for $n_2$) corresponding to the lowest row sums are removed (see \cref{fig:firstmethod} (Step 1)). \revision{When multiple leaves share the same value in the summation column, a tie-breaking rule is applied. The leaves are selected (trimmed) according to the numerical order of their labels.}

After trimming, we compute the distance matrices $D_1$ and $D_2$ for the remaining vertices in both trees. Each element $D_1(i, j)$ represents the distance between vertices $v_i$ and, $v_j$ in the first tree, $D_2$ is defined similarly for the second tree. These matrices are symmetric and of equal size (see \cref{fig:firstmethod} (Step 2)).

We then construct a complete bipartite graph $K_{n_r,n_r}$, where $n_r = \min\{n_1, n_2\}$  and  each edge $e_{ij}$ is assigned a weight defined as
\[
w_{ij} = \left|\|D_1(i)\|_2 - \|D_2(j)\|_2\right|,
\]
which approximates the $L_2$ distance between rows $i$ and $j$ of the respective matrices, i.e., $\|D_1(i) - D_2(j)\|_2 \sim \left|\|D_1(i)\|_2 - \|D_2(j)\|_2\right|$, while disregarding the order of vertices (see \cref{fig:firstmethod} (Step 3)).

Since both parts of the bipartite graph contain the same number of vertices (specifically, $n_r$), the minimum-weight maximum matching will consist of $n_r$ edges. Using the Hungarian algorithm, the time complexity of this step is $\min\{n_1, n_2\}^3$. In contrast, the approach in \cite{ASAoLMTfUV} has a running time of $\min\{n_1, n_2\}^2 \times \max\{n_1, n_2\}$ for this step. Thus, our method achieves a slight improvement in the running time for computing the minimum-weight maximum matching.


\revision{
Using the obtained matching, we assign labels to the trees and construct the induced matrices \( M_1 \) and \( M_2 \) (Step~4 in \cref{fig:firstmethod}). 
We then set
\[
\varepsilon = \| M_1 - M_2 \|_\infty
\]
and estimate the interleaving distance as
\[ \max\left\{ \frac{1}{2} \max_i \delta_i, \, \varepsilon \right\},\]
where \(\delta_i\) is the smallest value in the \(i\)-th row of the matrix \(S\), corresponding to an unknown trimmed vertex, after removing the trimmed vertices from the column set.
}

The method for the \emph{partial agreement} case is analogous. The number of known-labeled vertices is the same in both trees, and we trim the unknown-labeled vertices so that each tree contains an equal number of unknown-labeled vertices (denoted $n'_r = \min\{n'_1,n'_2\}$ where $n'_1$ is the number of unknownn leaves in $\tfp$ and $n'_2$ is the number of unknown leaves in $\tgp$). To trim the larger tree, as in the disagreement case, we construct the matrix $S$, where each row corresponds to an unknown-labeled vertex and each column corresponds to a leaf.

\begin{figure}[!ht]
\centering
\includegraphics[width=7cm, trim=20 290 300 20, clip]
{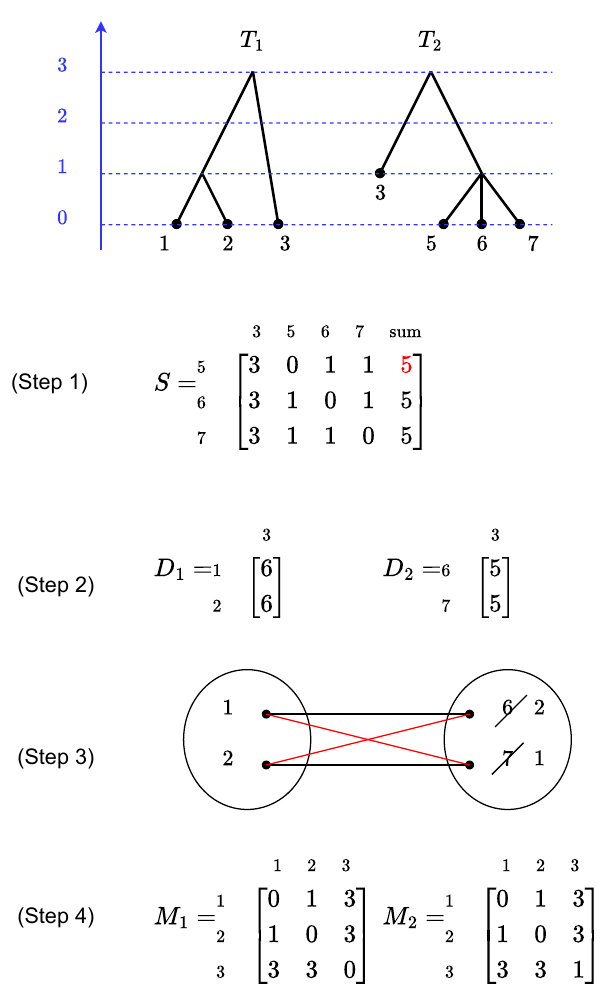}
\caption{Two labeled merge trees, $\tfp$ and $\tgp$, are illustrated, where the trees share one common label (label~3), representing a case of partial agreement. $\tfp$ is treated as the reference tree. In Step~1, we construct the matrix $S$ to identify \revision{the leaves} for trimming; in this example, the leaf with label~5 is trimmed (all rows sum to~5, so one is selected at random). In Step~2, we compute the distance matrices, which capture the distances between unknown-labeled vertices after trimming. In Step~3, we construct the bipartite graph and apply the minimum-weight maximum matching algorithm, resulting in labels 6 and 7 being matched to 2 and 1, respectively. Finally, in Step~4, we construct the induced matrices.}
\label{fig:firstmethod_2}
\end{figure}

We then remove $n'_1 - n'_2$ leaves corresponding to the rows of $S$ with the smallest row sums. \revision{If multiple leaves have the same value in the summation column, we select (or trim) them according to the numerical order of their labels.} (see Step~1 in \cref{fig:firstmethod_2} for illustration).

The main difference lies in the construction of the distance matrix \( D \). Following the approach in~\cite{ASAoLMTfUV}, as illustrated in \cref{sec:LabeledMergeTree}, we compute the distances from each unknown-labeled vertex to the known-labeled vertices. After trimming, both trees yield distance matrices \( D_1 \) and \( D_2 \) of equal dimensions, with $n_r$ rows and a number of columns equal to the number of known-labeled vertices. Here, \( D_1(i,j) \) denotes the distance from the unknown-labeled vertex \( v_i \) to the known-labeled vertex \( v_j \) in \( \tfp \); \( D_2(i,j) \) is defined similarly for \( \tgp \) (see \cref{fig:firstmethod_2} Step~2).

As in the disagreement case, we construct a complete bipartite graph \( K_{{n'_r},{n'_r}} \), where the vertices represent the trimmed unknown-labeled vertices. The weight of each edge is given by \( \| D_1(i) - D_2(j) \|_2 \). We then apply the minimum-weight maximum matching algorithm to determine the optimal pairing of unknown-labeled vertices (see Step~3 in \cref{fig:firstmethod_2}).


\revision{
After the matching phase, we generate the \(n_r \times n_r\) induced matrices \( M_1 \) and \( M_2 \) for the two trees (see Step~4 in \cref{fig:firstmethod_2}). 
The distance between \(T_1\) and \(T_2\) is defined as
\[\max\left\{ \frac{1}{2} \max_i \delta_i, \, \varepsilon \right\},\]
where \(\delta_i\) represents the minimum entry in the \(i\)-th row of \(S\) that corresponds to an unmatched trimmed vertex, after columns containing trimmed vertices have been removed. 
Here, \(\varepsilon\) is given by the infinity norm of \(M_1 - M_2\).
} \revision{ In \cref{alg:heuristicfirstmethod} in \cref{appen:appendixA}, the pseudo-code for the efficient leaf matching method for the partial agreement case is presented, for disagreement the code is similar.}

\begin{figure*}[t]
  \centering
  \subfloat[Example 1]{
        \includegraphics[height=2.4cm, trim=20 510 20 20, clip]{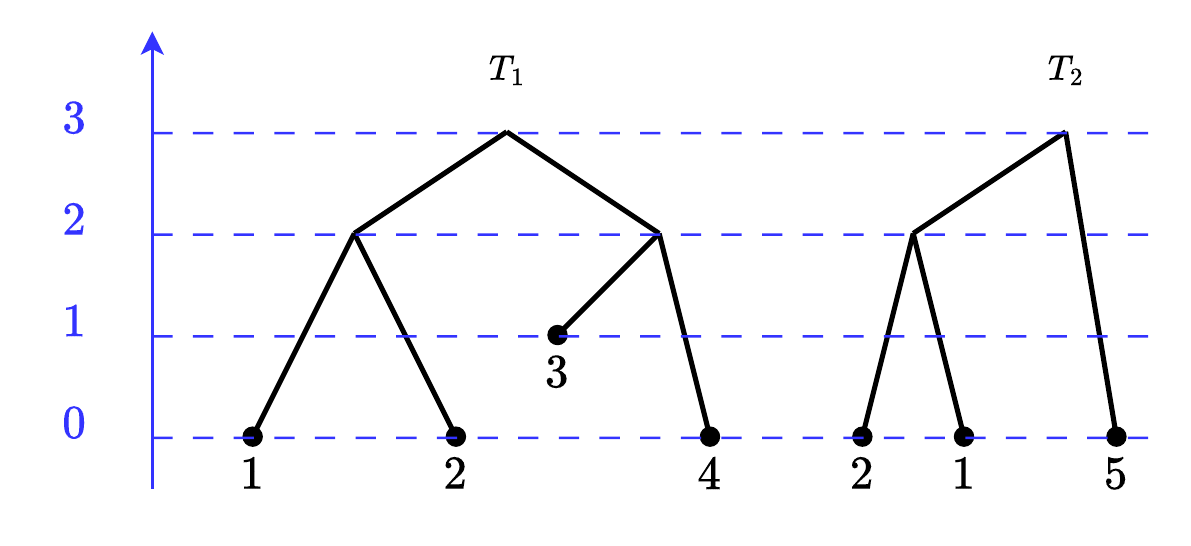} 
        \label{fig:toy1}

    }
    \subfloat[Example 2]{
        \includegraphics[height=4.1cm, trim=20 415 30 20, clip]{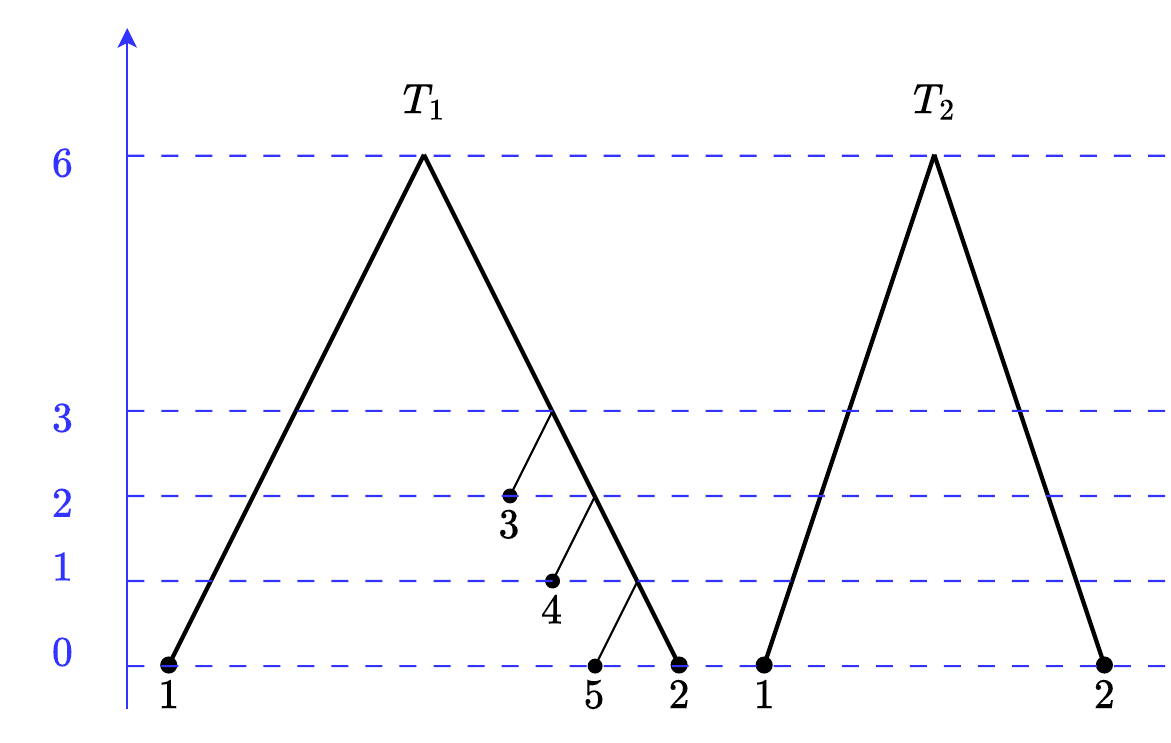}
        \label{fig:secondtoyexample}

    }
    \subfloat[Example 3]{
        \includegraphics[height=2.4cm, trim=50 490 20 20, clip]{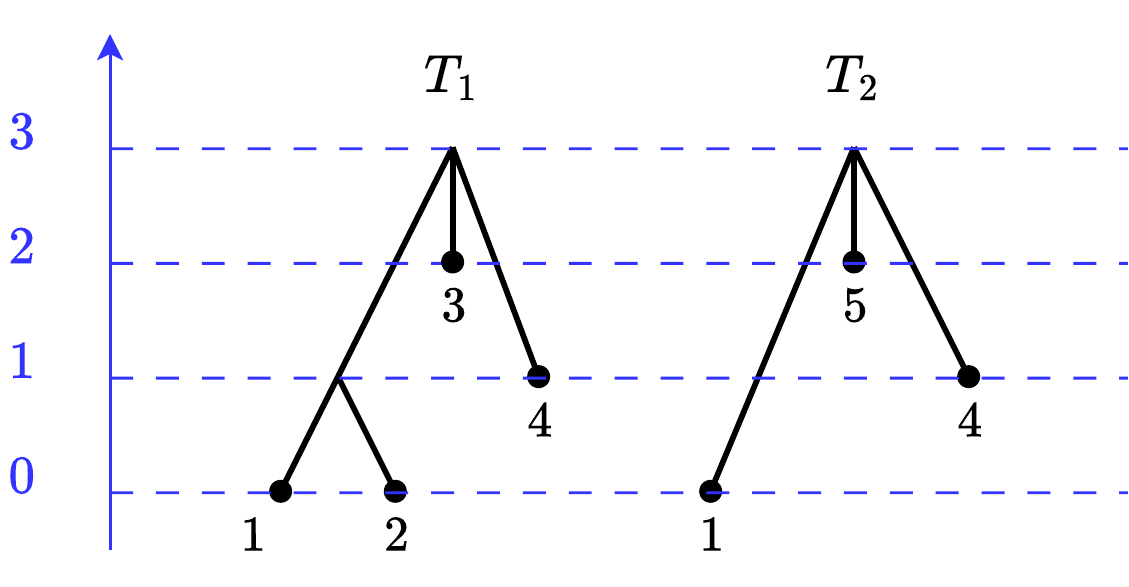}
        \label{fig:thirdtoyexample}

    }

  \caption{(a) An example of two labeled merge trees with a true interleaving distance of $0.5$, where the method in \cite{ASAoLMTfUV} overestimates the distance as $2$. (b) An example of partial agreement, where the true interleaving distance is $0.5$ and our first and second algorithms return the distance $0.5$. However, the method in \cite{ASAoLMTfUV} returns $3$.
  (c) An example of partial agreement, where the true interleaving distance is $0.5$ and our first method returns $2$. Besides, the method in \cite{ASAoLMTfUV} returns $1$. However, the second method return $0.5$.}
  \label{fig:examples}
\end{figure*}

\subsection{Modified Matching-Based Method (\MethodTwo)}

In this method, we make a slight modification to the approach proposed in~\cite{ASAoLMTfUV}, as described in \cref{sec:LabeledMergeTree}. Specifically, we adopt the same strategy as in the full agreement case. For partial agreement, we first construct a distance matrix that captures the distance between each vertex with an unknown label and all vertices with known labels. These matrices have the same number of columns.  In the next step, we build a bipartite graph in which the weight of each edge is given by the 2-norm of the difference between corresponding rows of the distance matrices. We then apply the minimum-weight maximum matching algorithm to the bipartite graph. Once a matching of size $\min\{n'_1, n'_2\}$ is obtained---where $n'_1$ and $n'_2$ denote the number of vertices in the two partitions of the bipartite graph---we compute $\delta_i$ for the remaining unmatched leaves in the pivot tree (i.e., the tree with more unknown labels) using the method described in \cref{sec:FasterAlgorithm}. To do this, we construct the matrix $S$ for the pivot tree. The rows of $S$ correspond to the unmatched leaves, and the columns correspond to all other vertices (excluding the vertex of that row). For each row, $\delta_i$ is defined as the minimum value in that row. 
For each row, $\delta_i$ is defined as the minimum entry in that row. The distance between the two trees is computed as
\[
\max\left\{ \frac{1}{2} \max_i \delta_i, \, \varepsilon \right\},
\]
where $\varepsilon$ denotes the infinity norm of the difference between the induced matrices.


The disagreement case follows a similar procedure. However, in this case, the weighted bipartite graph is constructed using edge weights of the form $\|D_1(i) - D_2(j)\|_2$, where $D_1(i)$ denotes the $i$-th row of the distance matrix of the first tree, and $D_2(j)$ is defined analogously for the second tree. \revision{The algorithm for the modified matching-based method for the partial agreement case is presented in \cref{alg:heuristicsecondmethod} in \cref{appen:appendixB}.}

\subsection{Illustrative \revision{Examples}}

\revision{\textbf{First Example.}}
We consider two merge trees, $\tfp$ and $\tgp$, as presented in \cite{ASAoLMTfUV} and reproduced here in \cref{fig:toy1}. These two trees have an interleaving distance of $0.5$. Here, $\tgp$ is the reference tree, and using our first method, we construct a non-symmetric matrix $S$ for the tree with more unknown labels, which in this case is $\tfp$, as follows:
\[
S = 
\begin{array}{c@{}c}
  & 
  \begin{array}{ccccc}
    ~~~\scriptstyle 1 & 
    \scriptstyle 2 & 
    \scriptstyle 3 &
    \scriptstyle 4 &
    \scriptstyle \text{sum}
  \end{array} \\

  \begin{array}{c}
    \scriptstyle 3 \\
    \scriptstyle 4 
  \end{array} &
  \left[
    \begin{array}{ccccc}
        2 & 2 & 0 & 1 &  {\color{red}5}\\
        3 & 3 & 2 & 0 &  8
    \end{array}
  \right]
\end{array}
\]

The red number in the last column of $S$ indicates the minimum value in that column and therefore indicates the label of the row that should be trimmed — in this case, the vertex labeled $3$. After trimming, each tree has only one vertex remaining. We then construct a bipartite graph, which results in a single edge used in the minimum-weight maximum matching. As a result, we rename vertex $4$ in tree $\tfp$ to $5$.

In the next step, we construct two matrices, $M_1$ and $M_2$, which are identical in this case, as follows:

\[
M_1 = M_2 = 
\begin{array}{c@{}c}
  & 
  \begin{array}{ccc}
    \scriptstyle 1 & 
    \scriptstyle 2 & 
    \scriptstyle 5 
  \end{array} \\

  \begin{array}{c}
    \scriptstyle 1 \\
    \scriptstyle 2 \\
    \scriptstyle 5
  \end{array} &
  \left[
    \begin{array}{ccc}
        0 & 2 & 3 \\
        2 & 0 & 3 \\
        3 & 3 & 0        
    \end{array}
  \right]
\end{array}
\]
To compute $\delta$ we consider the following matrix: 
\[
S = 
\begin{array}{c@{}c}
  & 
  \begin{array}{ccccc}
    \scriptstyle 1 & 
    \scriptstyle 2 & 
    \scriptstyle 3 &
    \scriptstyle 4 
  \end{array} \\

  \begin{array}{c}
    \scriptstyle 3 \\
    \scriptstyle 4 
  \end{array} &
  \left[
    \begin{array}{ccccc}
        2 & 2 & \cancel{0} & {\textcolor{red}{1}} \\
        3 & 3 & \cancel{2} & 0 
    \end{array}
  \right]
\end{array}
\]
$\delta$ is calculated to be $1$.
The distance between the trees is computed as:

\[
\max\left\{\frac{\delta
}{2}, \|M_1 - M_2\|_\infty\right\} = 0.5
\]
which matches the interleaving distance. 

Using our second method, we construct two matrices 
\[
U_p = 
\begin{array}{c@{}c}
  & 
  \begin{array}{cc}
    \scriptstyle 1 & 
    \scriptstyle 2
  \end{array} \\

  \begin{array}{c}
    \scriptstyle 3 \\
    \scriptstyle 4 
  \end{array} &
  \left[
    \begin{array}{cc}
        5 & 5  \\
        6 & 6
    \end{array}
  \right]
\end{array}, \quad
U_2 = 
\begin{array}{c@{}c}
  & 
  \begin{array}{cc}
    \scriptstyle 1 & 
    \scriptstyle 2
  \end{array} \\

  \begin{array}{c}
    \scriptstyle 5 
  \end{array} &
  \left[
    \begin{array}{cc}
        6 & 6
    \end{array}
  \right].
\end{array}
\]
By constructing the bipartite graph, leaf $4$ is matched with leaf $5$. Using the $S$ matrix, we obtain the same result as with the first method.

In contrast, applying the method from \cite{ASAoLMTfUV} yields a distance of $2$, which deviates significantly from the correct interleaving distance.

\revision{\textbf{Second Example.} We now consider a second illustrative example, shown in \cref{fig:secondtoyexample}, which corresponds to the partial agreement case depicted in \cref{fig:disagreement}. In this scenario, leaves labeled $3$, $4$, and $5$ in tree $T_1^{f,\pi_1}$ are unknown-labeled, while leaves labeled $1$ and $2$ are known-labeled in both trees.  
}

\revision{
Using our first method, all unknown-labeled leaves are trimmed, yielding:}
\[
M_1 = M_2 = 
\begin{array}{c@{}c}
  & 
  \begin{array}{cc}
    \scriptstyle 1 & 
    \scriptstyle 2 
  \end{array} \\

  \begin{array}{c}
    \scriptstyle 1 \\
    \scriptstyle 2 
  \end{array} &
  \left[
    \begin{array}{cc}
        0 & 6 \\
        6 & 0        
    \end{array}
  \right]
\end{array}
\]
\revision{
For all unknown-labeled leaves, we have $\delta_i = 1$, giving an estimated distance of $0.5$. Our second method yields the same result. }

\revision{
In contrast, applying the method from \cite{ASAoLMTfUV} assigns the leaf labeled $2$ additional labels $3$, $4$, and $5$, which increases the computed distance to $3$.
}

\revision{\textbf{Third Example.} In this example, we consider the two labeled merge trees shown in \cref{fig:thirdtoyexample}. The $S$ matrix is constructed for the larger tree, $T_1^{f,\pi_1}$, as follows:}

\[
S = 
\begin{array}{c@{}c}
  & 
  \begin{array}{ccccc}
    ~~~\scriptstyle 1 & 
    \scriptstyle 2 & 
    \scriptstyle 3 &
    \scriptstyle 4 &
    \scriptstyle \text{sum}
  \end{array} \\

  \begin{array}{c}
    \scriptstyle 2 \\
    \scriptstyle 3 
  \end{array} &
  \left[
    \begin{array}{ccccc}
        1 & 0 & 3 & 3 &  {7}\\
        1 & 1 & 0 & 1 &  {\color{red} 3}
    \end{array}
  \right]
\end{array}.
\]

\revision{The leaf with label $3$ is selected as a candidate for trimming. After trimming, both trees have only one unlabeled leaf, which are matched to each other using the bipartite graph and the minimum-weight maximum matching. Therefore, the induced matrices are as follows: }
\[
M_1 = 
\begin{array}{c@{}c}
  & 
  \begin{array}{ccc}
    \scriptstyle 1 & 
    \scriptstyle 2 & 
    \scriptstyle 4 
  \end{array} \\

  \begin{array}{c}
    \scriptstyle 1 \\
    \scriptstyle 2 \\
    \scriptstyle 4
  \end{array} &
  \left[
    \begin{array}{ccc}
        0 & 1 & 3 \\
        1 & 0 & 3 \\
        3 & 3 & 1        
    \end{array}
  \right]
\end{array},  
\quad
M_2 = 
\begin{array}{c@{}c}
  & 
  \begin{array}{ccc}
    \scriptstyle 1 & 
    \scriptstyle 2 & 
    \scriptstyle 4 
  \end{array} \\

  \begin{array}{c}
    \scriptstyle 1 \\
    \scriptstyle 2 \\
    \scriptstyle 4
  \end{array} &
  \left[
    \begin{array}{ccc}
        0 & 3 & 3 \\
        3 & 2 & 3 \\
        3 & 3 & 1        
    \end{array}
  \right]
\end{array}
\]

\revision{The estimated distance using the first method is $2$. However, by applying the greedy method described in \cite{ASAoLMTfUV}, we take the larger tree as the pivot tree. In $T_2$, after constructing the bipartite graph and solving the minimum-weight maximum matching problem, the leaf labeled~$1$ is additionally assigned the label~$2$. By calculating the induced matrices, the distance is estimated as $1$. In contrast, our second method returns $0.5$, which corresponds to the true interleaving distance between the two given merge trees.}

\revision{In \cref{fig:exampleall}, we compare all three methods using a single illustrative example.}

\section{Experimental Results}

\begin{figure*}[t]
  \centering
  \includegraphics[trim={0 3cm 11.8cm 2cm},clip,height=4.4cm]{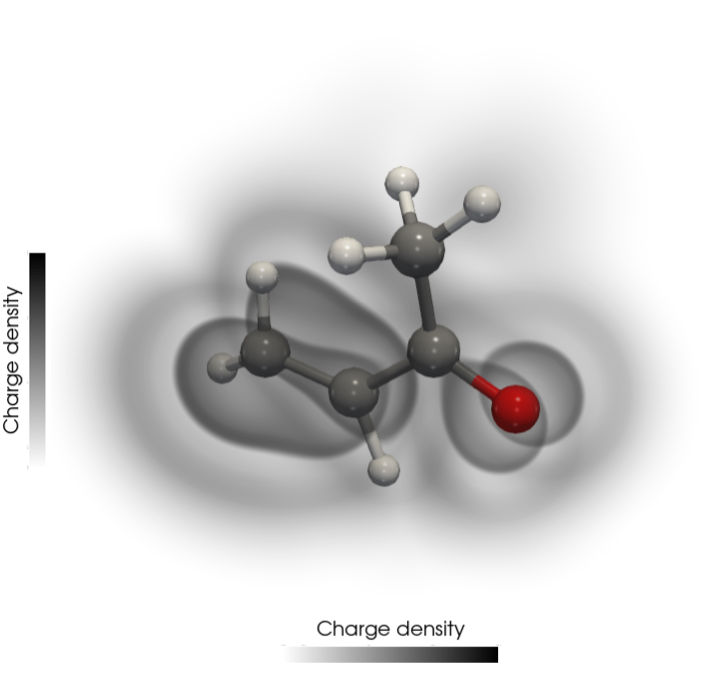}
  \subfloat[MVK molecule]{
        \includegraphics[trim={0 4cm 0 2cm},clip,height=4.4cm]{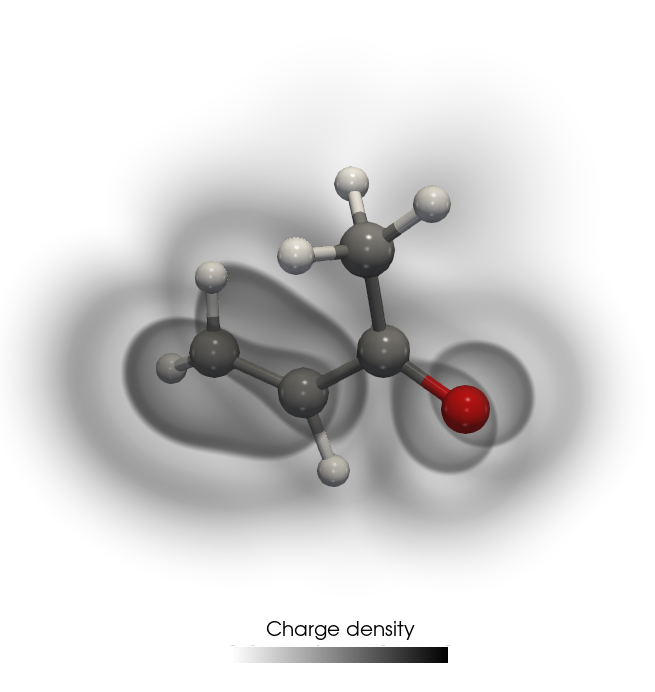}
        \label{subfig:scalar_field}

    }
    \subfloat[Critical points]{
        \includegraphics[trim={0 4cm 0 2cm},clip,height=4.4cm]{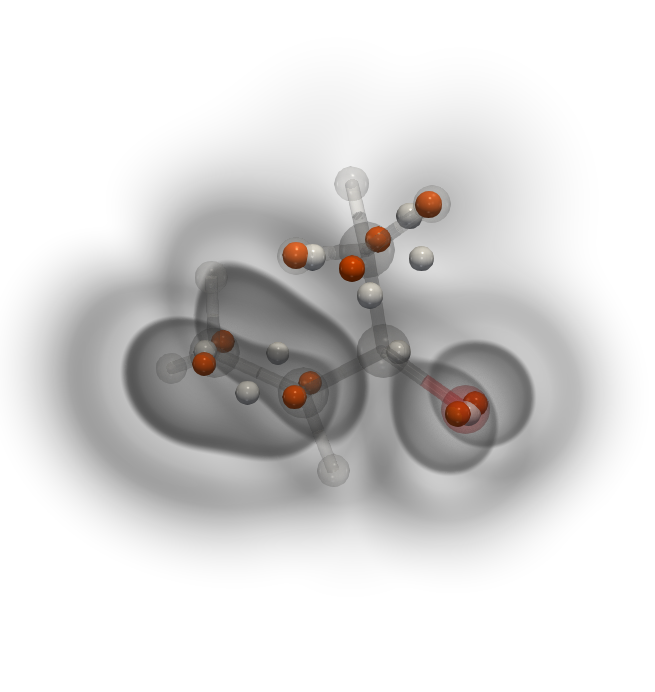}
        \label{subfig:cps}

    }
    \subfloat[Labeled merge tree]{
        \includegraphics[trim={0 4cm 0 2cm},clip,height=4.4cm]{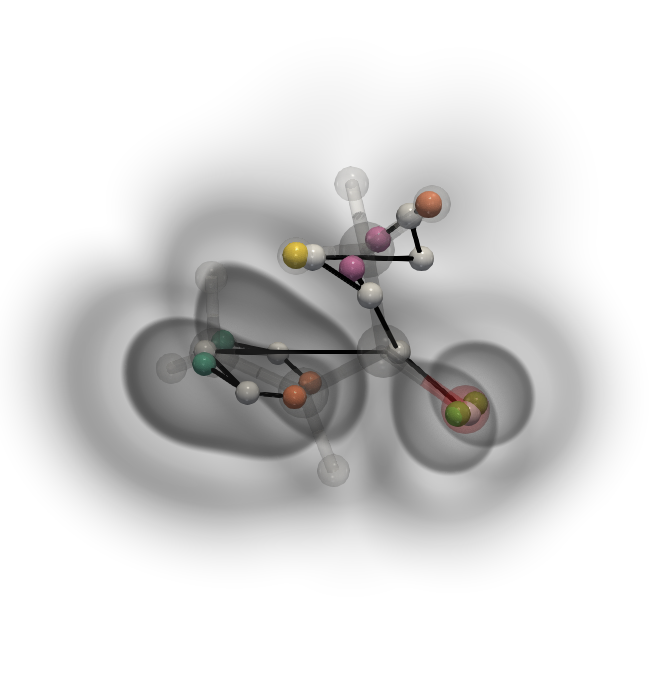}
        \label{subfig:mt}

    }

    \caption{One time step selected from \StateTwoP is shown as an example. For each time step, the data set contains information about atom locations and the electron density distribution. (a) The atoms are shown as spheres colored according to atom type along with volume rendering of the charge density field. (b) The maxima (orange) and saddles (white) are shown as smaller spheres. (c) A labeling can be assigned based on Voronoi segmentation induced by atom locations, resulting in a labeled merge tree for each time step. \revision{The white spheres correspond to the saddles while different colors are used to indicate different labels}.}
    
  \label{fig:MVK}
\end{figure*}

\begin{figure*}[!ht]
    \raggedright
    \hspace{1.2em}
    \includegraphics[width=0.2\linewidth]{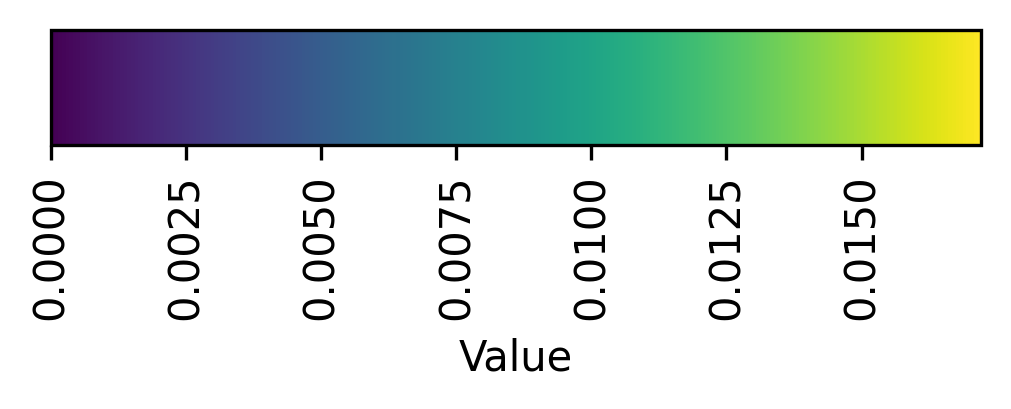} 
    \vspace{0.5em} 
    
    \centering

    \makebox[\linewidth][c]{%
        \hspace{-3em} 
        \begin{minipage}{0.8\linewidth}
            \centering
            \begin{tabular}{ccccc}
                \textbf{\MethodOne~~~~\qquad\qquad} & 
                \textbf{\MethodTwo ~~\qquad\qquad } & 
                \textbf{ Greedy~\cite{ASAoLMTfUV}\qquad\qquad} & 
                \textbf{ M1 vs Greedy\qquad\qquad} & 
                \textbf{M2 vs Greedy }
            \end{tabular}
        \end{minipage}
    }

    \vspace{0.5 em}

    \begin{tabular}{@{}c@{}c@{}}
    \raisebox{2cm}{\rotatebox[origin=c]{90}{\textbf{\StateOneH}}} &
    \includegraphics[width=0.95\linewidth]{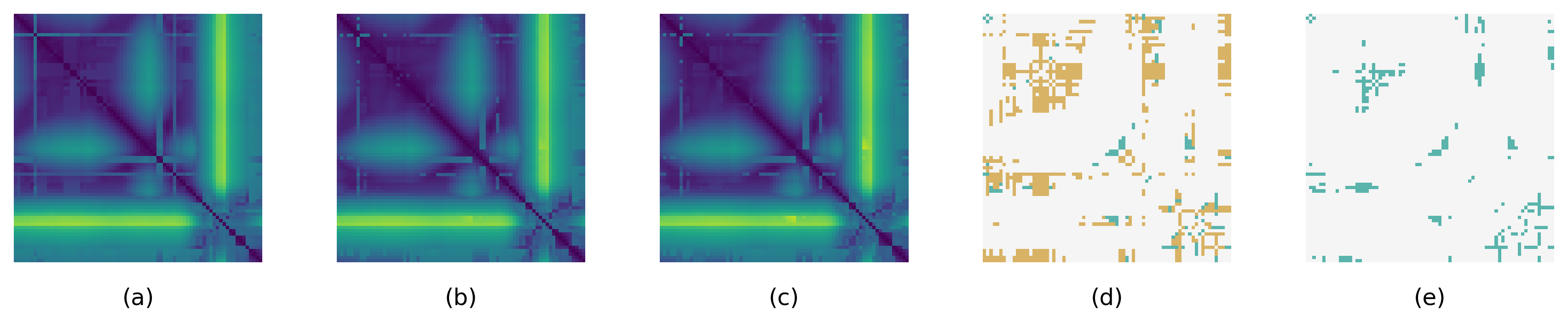} \\

    \raisebox{2cm}{\rotatebox[origin=c]{90}{\textbf{\StateOneP}}} &
    \includegraphics[width=0.95\linewidth]{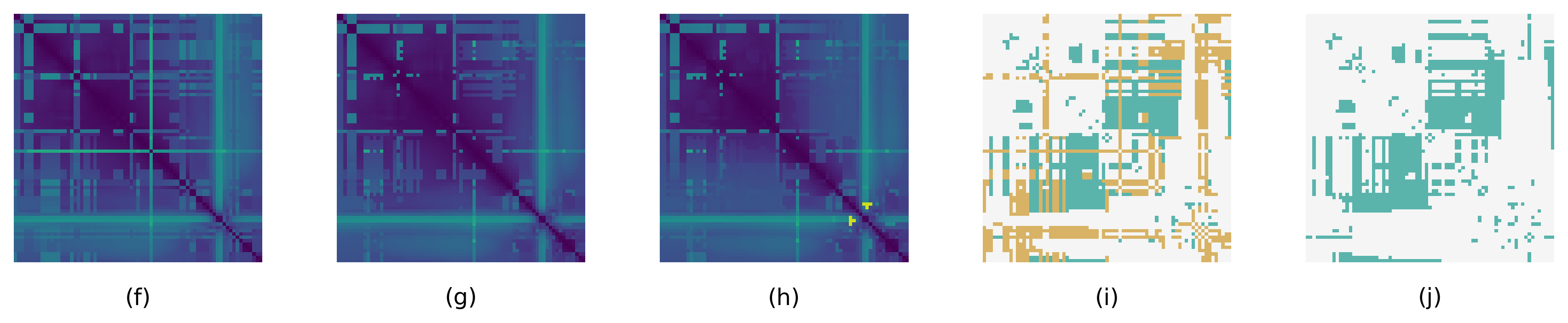} \\

    \raisebox{2cm}{\rotatebox[origin=c]{90}{\textbf{\StateTwoH}}} &
    \includegraphics[width=0.95\linewidth]{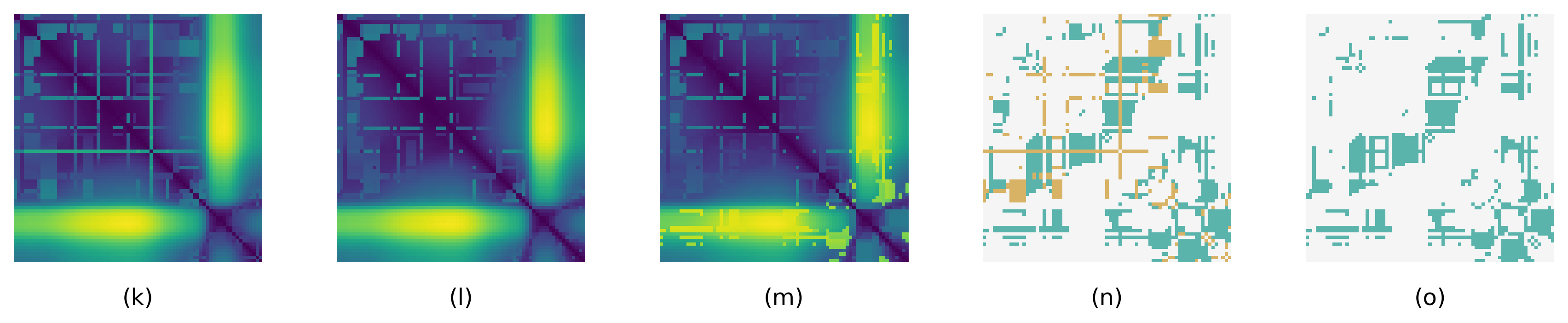} \\

    \raisebox{2cm}{\rotatebox[origin=c]{90}{\textbf{\StateTwoP}}} &
    \includegraphics[width=0.95\linewidth]{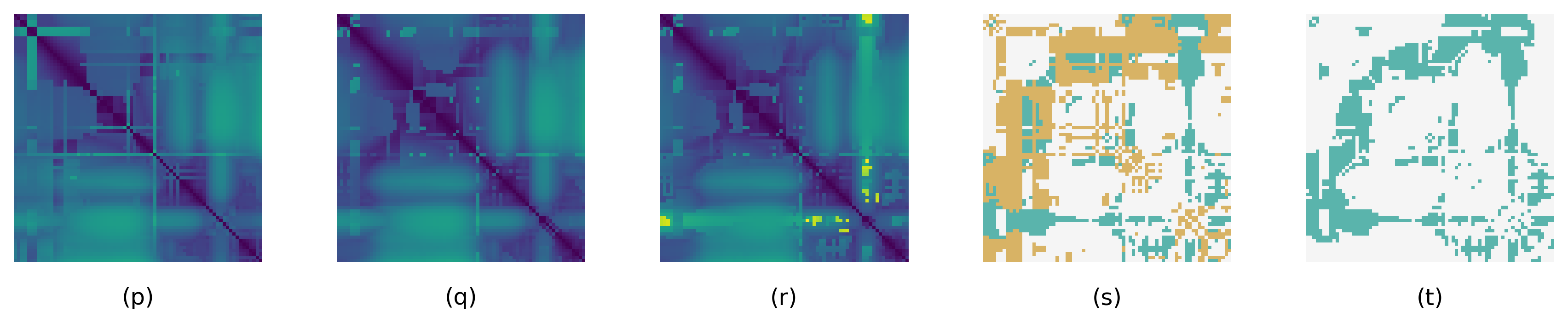} \\
    \end{tabular}

   \caption{$75\times75$ distance matrices for the datasets \StateOneH, \StateOneP, \StateTwoH, and \StateTwoP. The first column—(a), (f), (k), and (p)—shows the distance matrices computed using our first heuristic method. The second column—(b), (g), (l), and (q)—presents the corresponding matrices from our second method. The third column—(c), (h), (m), and (r)—displays the results from the previous greedy method~\cite{ASAoLMTfUV}. The fourth column illustrates the comparative performance between our first heuristic (first column) and the greedy method (third column): blue regions indicate where our heuristic performs better, yellow regions where the greedy method outperforms, and gray regions where both methods yield identical results. The final column compares our second method (second column) with the greedy method (third column): blue areas show where our second method performs better, and gray areas indicate equal performance.}
    \label{fig:real}
    \vspace{-1em}
\end{figure*}

\begin{table*}[!htbp]
\centering
\resizebox{\textwidth}{!}{%
\begin{tabular}{
    l
    c
    c
    c
    c
    c
    c
    c
    c
    c
}
\toprule
& & & \multicolumn{3}{c}{\textbf{Running time (sec)}} & \multicolumn{2}{c}{\textbf{Greedy vs. \MethodOne}} & \multicolumn{2}{c}{\textbf{Greedy vs. \MethodTwo}} \\
\cmidrule(lr){4-6} \cmidrule(lr){7-8} \cmidrule(lr){9-10}
\textbf{Dataset} &
\textbf{Avg. Vertices} &
\textbf{Avg. $|u_1 - u_2|$} &
\textbf{\MethodOne} &
\textbf{\MethodTwo} &
\textbf{Greedy Method}&
\textbf{$G>M1$} &
\textbf{$M1>G$} &
\textbf{$G>M2$} & 
\textbf{$M2>G$} \\
\midrule
\texttt{Random\_50}  & $40.0$ & $4.83$ & $0.91$ & $0.91$ & $1.18$ & $24.50$ & $42.00$ & $40.50$ & $0.0$ \\
\texttt{Random\_100}  & $82.9$ & $7.98$ & $3.96$ & $4.08$ & $5.18$ & $11.00$ & $57.50$ & $43.00$ & $0.0$  \\
\texttt{Random\_200}  & $154.5$ & $24.75$ & $17.82$ & $19.66$ & $25.32$ & $14.50$ & $61.50$ & $47.00$ & $0.0$  \\
\texttt{Random\_500}  & $405.0$ &$52.19$ & $237.81$ & $279.30$ & $320.61$ & $28.50$ &$50.50$ & $61.50$ & $0.0$ \\
\midrule
\StateOneH  & 27.05 & 1.43 & $7.49$ & $7.39$ & $8.70$ &$2.03$ & $14.12$& $4.76$ & $0.0$ \\
\StateOneP  & 27.55 & 2.20 & $7.59$ & $7.49$ & $9.27$ & $18.03$& $20.05$& $23.54$ & $0.0$ \\
\StateTwoH  & 26.32 & 1.90 & $7.10$ & $6.99$ & $8.50$ & $19.66$ & $7.15$ & $19.70$ & $0.0$ \\
\StateTwoP   & 29.63 & 2.89 & $8.62$ & $8.56$ & $10.76$ &$16.82$& $29.65$& $26.74$ & $0.0$ \\
\bottomrule
\end{tabular}%
}
\caption{\revision{Comparison of average properties and run-times across different datasets.  
\emph{Avg. Vertices} denotes the average total number of vertices. 
\emph{Avg. $|u_1 - u_2|$} indicates the average absolute difference in the number of unknown labels between trees. 
\MethodOne and \MethodTwo refer to the two heuristic methods proposed in this paper. 
\emph{Greedy Method} refers the approach described in~\cite{ASAoLMTfUV} for computing the distance matrix. The running times are reported for computation of the complete distance matrices. The distance matrix for synthetic datasets contain $^{20}C_2$ entries while for the chemistry dataset there are $^{20}C_2$ entries.
}
}
\label{tab:table1}
\end{table*}

\revision{
We evaluate the performance of our methods against the greedy approach described in \cite{ASAoLMTfUV}, using both synthetic and real-world ensembles of partially labeled merge trees. The first metric considered is running time. As a second metric, we assess the quality of the computed tree distances produced by the three methods. Since computing the exact interleaving distance between two trees is NP-hard, we instead compare the number of cases in which our method yields a smaller distance value than the greedy approach of \cite{ASAoLMTfUV}. A smaller distance value is preferable, as it indicates a result closer to the ground truth.
}
\subsection{Synthetic Data}
\label{sec:synthetic}
\revision{
We generate a synthetic ensemble of partially labeled merge trees to evaluate the scalability of the methods on large trees. Given a user-defined maximum number of vertices, a random \emph{base merge tree} is constructed. This is achieved by first generating a random full binary tree, where each node has either two or zero children. Starting from the root node, the tree is expanded to the specified size by iteratively attaching two vertices to a randomly chosen leaf at each step. Each edge in the tree is then assigned a random length uniformly sampled from the interval $[0,1]$. The scalar value of each node is defined as the negative distance from the root. Finally, a user-specified proportion of the leaves in the resulting merge tree are assigned labels. In our experiments, this parameter is set to $0.5$, meaning that half of the leaves are labeled while the remaining leaves remain unlabeled.}

\revision{
From this partially labeled base merge tree, we generate an ensemble of perturbed merge trees. Three types of perturbations are applied in sequence: scalar value updates, tree rotations, and leaf deletions. In the scalar value update step, a random subset of vertices is selected and their scalar values are perturbed to new values. Next, a random number of tree rotations are applied to selected internal vertices, thereby altering the structure of the merge tree. Finally, a random number of leaves are deleted to introduce variation in tree size across the ensemble.}

\revision{
We generate four base merge trees with sizes of 50, 100, 200, and 500 vertices. For each base tree, we construct an ensemble of twenty perturbed merge trees with progressively increasing perturbation. We refer to these synthetic ensembles as \texttt{Random\_50}, \texttt{Random\_100}, \texttt{Random\_200}, and \texttt{Random\_500}, respectively.}

\subsection{A Dataset from Chemistry}
\label{sec:real}
\revision{Next, we consider} a simulation dataset from the domain of theoretical photochemistry, representing light-induced dynamics of methylvinylketone (MVK), a molecule present in lower atmosphere~\cite{chakraborty2023}. The dataset captures the coupled evolution of electronic structure and nuclear geometry over time following photoexcitation. The electronic structure, relevant for studying excited-state transitions, is described using the hole and particle natural transition orbitals (NTOs), which represent the spatial distribution of electron depletion and accumulation, respectively~\cite{martin_nto_2023}. The dataset focuses on the early-stage dynamics (first 36 femtoseconds) involving two electronic excited states, S1 and S2 sampled over $75$ time steps. For each state, the hole and particle NTOs are provided as time-varying electronic density fields, resulting in a total of four dynamic volumetric scalar fields. Additionally, the trajectory includes the time-resolved atomic positions of all $11$ atoms comprising the MVK molecule, allowing for correlated analysis of nuclear and electronic evolution.
%
Wetzels et al.~\cite{EEDEUMTM,wetzels2025accelerating} used merge-tree-based edit distance metrics to analyze the evolution of the S1 electronic state in the MVK dataset. Their work focused on identifying correlations between variations in merge tree structures and changes in interatomic distances, with the goal of revealing patterns in the joint evolution of electronic and nuclear configurations. More recently, Sharma et al.~\cite{sharma2025continuous} employed bivariate analysis based on the peeling of continuous scatter plots for the analysis of the same dataset. 

In contrast to these previous approaches, we introduce a novel analysis based on labeled merge trees. Refer to \cref{fig:MVK} for an illustrative example. Specifically, we assign semantic labels to the vertices of each merge tree by leveraging a Voronoi segmentation of the spatial domain, where the regions are defined by proximity to the nuclei of the atoms of the molecule. Each local maximum in the electron density field is assigned the label of its nearest atom; if multiple maxima are assigned the same label, we retain the label only for the maximum with the highest persistence, treating other maxima within the same Voronoi region as unlabeled. \revision{Two example merge trees are shown in \cref{fig:trees} of the Appendix.}
Using this labeling strategy, we construct four independent time series of \revision{partially} labeled merge trees, corresponding to the four time-varying scalar fields derived from the hole and particle NTOs in states S1 and S2. We refer to these four labeled merge tree sequences as \StateOneH, \StateOneP, \StateTwoH, and \StateTwoP, respectively.

\subsection{Observations and Discussion}
\revision{\cref{tab:table1} summarizes the results of our experiments on the two datasets described above. All experiments were conducted on a MacBook Pro with an Apple M4 chip, 16\,GB RAM, running macOS Sequoia 15.5. For each ensemble, we generated the distance matrix for each of the three methods, recording the interleaving distance between each pair of trees in the matrix. Running times correspond to the total time needed to create this distance matrix for each method averaged over 10 independent runs. We consistently observe that both our methods are faster than the greedy approach suggested in \cite{ASAoLMTfUV}, with \MethodOne outperforming \MethodTwo.}   

\revision{
Next we analyze and compare the quality of the distance values computed. \cref{fig:real} shows a detailed comparison of the distance matrices obtained by the three methods for the chemistry dataset described in \cref{sec:real}. For synthetic tree ensembles described in \cref{sec:synthetic}, similar figure is provided in the appendix (\cref{fig:random}). A general observation is that while \MethodOne is faster than the greedy method \cite{ASAoLMTfUV}, it can produce worse results in some cases, that is, the distance value obtained is greater than that produced by the greedy method. However, we did not observe any case where \MethodTwo performs worse than the greedy method \cite{ASAoLMTfUV}. 
}
%
%

Furthermore, when analyzing each dataset individually, we observe that an increase in the average value of $U = |u_1 - u_2|$, where $u_1$ and $u_2$ are the unknown labels from the first and second trees, respectively, correlates with an increase in running time. (see \cref{tab:table1}.)
In addition, we observe that the performance of the previous method degrades more significantly than that of our first proposed method. 


\section{Conclusion}


\revision{
In this paper, we introduced two efficient heuristic algorithms for approximating the interleaving distance between labeled merge trees, a fundamental problem in topological data analysis. These methods are designed to address the limitations of the previously proposed greedy algorithm by Yan et al. \cite{ASAoLMTfUV}.
}

\revision{
The first method, referred to as the efficient leaf matching method, offers reduced computational complexity and runtime. It is particularly effective for large trees or in scenarios with a significant imbalance between the number of known-labeled and unknown-labeled leaves. Despite its simplicity, this method achieves accuracy comparable to the previous approach for estimating the interleaving distance, as proposed in \cite{ASAoLMTfUV}, in most cases.}

\revision{
The second method, referred to as the modified matching-based method, is designed as an improvement over the approach introduced in \cite{ASAoLMTfUV}. While it incurs a slightly higher computational cost compared to the first method, it consistently matches or exceeds the accuracy of the greedy method illustrated in \cite{ASAoLMTfUV}. This makes it a strong alternative for trees with fewer leaves or in scenarios where execution time is less critical.}

\revision{
Based on our empirical evaluation, we recommend using the first method for large merge trees, in disagreement scenarios, or when, in cases of partial agreement, the number of unknown-labeled leaves significantly exceeds the number of known-labeled ones (this method also yields better results than the second method in some cases). For smaller trees or those with a more balanced leaf distribution, the second method provides faster execution and greater accuracy.}

\revision{
In future work, we plan to extend these heuristic approaches to broader applications, such as solving the 1-center and average problems on labeled merge trees. These extensions would further enhance the practical utility of our methods in summarizing and comparing tree-structured data in various scientific and engineering domains.
}

\acknowledgments{
   We thank Dr. Nanna H. List (KTH Royal Institute of Technology, Stockholm) for providing the simulation dataset used for evaluation in this paper. This work was partially supported by the the Swedish Research Council (VR) grant 2023-04806; the Swedish e-Science Research Center (SeRC); and the Wallenberg Autonomous Systems and Software Program (WASP) funded by the Knut and Alice Wallenberg Foundation.
}

\bibliographystyle{abbrv-doi}

\bibliography{Reference}

\begin{thebibliography}{10}

\bibitem{CtGHDfMT}
P.~K. Agarwal, K.~Fox, A.~Nath, A.~Sidiropoulos, and Y.~Wang.
\newblock Computing the {{G}romov-{H}ausdorff} distance for metric trees.
\newblock {\em ACM Trans. Algorithms}, 14(2):24:1--24:20, 2018.

\bibitem{biasotti2008describing}
S.~Biasotti, L.~De~Floriani, B.~Falcidieno, P.~Frosini, D.~Giorgi, C.~Landi, L.~Papaleo, and M.~Spagnuolo.
\newblock Describing shapes by geometrical-topological properties of real functions.
\newblock {\em ACM Computing Surveys (CSUR)}, 40(4):1--87, 2008.

\bibitem{ASoTEDaRP}
P.~Bille.
\newblock A survey on tree edit distance and related problems.
\newblock {\em Theoretical Computer Science}, 337(1-3):217--239, 2005.

\bibitem{bondy}
J.~A. Bondy and U.~S.~R. Murty.
\newblock {\em Graph Theory with Applications}.
\newblock Elsevier, New York, 1976.

\bibitem{IC}
R.~A. Brualdi.
\newblock {\em Introductory Combinatorics}.
\newblock Pearson Prentice Hall, Saddle River, N.J., 4th ed., 2004.

\bibitem{carr2003computing}
H.~Carr, J.~Snoeyink, and U.~Axen.
\newblock Computing contour trees in all dimensions.
\newblock {\em Computational Geometry}, 24(2):75--94, 2003.

\bibitem{chakraborty2023}
P.~Chakraborty, R.~C. Couto, and N.~H. List.
\newblock Deciphering methylation effects on {S2($\pi \pi^*$)} internal conversion in the simplest linear {$\alpha,\beta$}-unsaturated carbonyl.
\newblock {\em The Journal of Physical Chemistry A}, 127(25):5360--5373, 2023.
\newblock PMID: 37331016. doi: {{%
10\hspace{.1pt}\discretionary{.}{%
}{.}\hspace{.4pt}1021\discretionary{/}{%
}{/}acs\hspace{.1pt}\discretionary{.}{%
}{.}\hspace{.4pt}jpca\hspace{.1pt}\discretionary{.}{%
}{.}\hspace{.4pt}3c02582}}


\bibitem{PoPMaTD}
F.~Chazal, D.~Cohen-Steiner, M.~Glisse, L.~J. Guibas, and S.~Y. Oudot.
\newblock Proximity of persistence modules and their diagrams.
\newblock In {\em Proceedings of the Twenty-Fifth Annual Symposium on Computational Geometry (SoCG)}, pp. 237--246, 2009. doi: {{%
10\hspace{.1pt}\discretionary{.}{%
}{.}\hspace{.4pt}1145\discretionary{/}{%
}{/}1542362\hspace{.1pt}\discretionary{.}{%
}{.}\hspace{.4pt}1542407}}


\bibitem{BToDL}
C.~Flamm, I.~L. Hofacker, P.~F. Stadler, and M.~T. Wolfinger.
\newblock Barrier trees of degenerate landscapes.
\newblock {\em Zeitschrift f{\"u}r Physikalische Chemie}, 216(2):155--173, 2002. doi: {{%
10\hspace{.1pt}\discretionary{.}{%
}{.}\hspace{.4pt}1524\discretionary{/}{%
}{/}zpch\hspace{.1pt}\discretionary{.}{%
}{.}\hspace{.4pt}2002\hspace{.1pt}\discretionary{.}{%
}{.}\hspace{.4pt}216\hspace{.1pt}\discretionary{.}{%
}{.}\hspace{.4pt}2\hspace{.1pt}\discretionary{.}{%
}{.}\hspace{.4pt}155}}


\bibitem{IIDfMT}
E.~Gasparovic, E.~Munch, S.~Oudot, K.~Turner, B.~Wang, and Y.~Wang.
\newblock Intrinsic interleaving distance for merge trees.
\newblock {\em La Matematica}, 4:40--65, 2025. doi: {{%
10\hspace{.1pt}\discretionary{.}{%
}{.}\hspace{.4pt}1007\discretionary{/}{%
}{/}s44007\discretionary{%
}{-}{-}024\discretionary{%
}{-}{-}00143\discretionary{%
}{-}{-}9}}


\bibitem{heine2016survey}
C.~Heine, H.~Leitte, M.~Hlawitschka, F.~Iuricich, L.~De~Floriani, G.~Scheuermann, H.~Hagen, and C.~Garth.
\newblock A survey of topology-based methods in visualization.
\newblock In {\em Computer Graphics Forum}, vol.~35, pp. 643--667. Wiley Online Library, 2016.

\bibitem{AoTAAtTE}
T.~Jiang, L.~Wang, and K.~Zhang.
\newblock Alignment of trees: An alternative to tree edit.
\newblock {\em Theoretical Computer Science}, 143(1):137--148, 1995.

\bibitem{lohfink2020fuzzy}
A.-P. Lohfink, F.~Wetzels, J.~Lukasczyk, G.~H. Weber, and C.~Garth.
\newblock Fuzzy contour trees: Alignment and joint layout of multiple contour trees.
\newblock In {\em Computer Graphics Forum}, vol.~39, pp. 343--355. Wiley Online Library, 2020.

\bibitem{martin_nto_2023}
R.~L. Martin.
\newblock {Natural transition orbitals}.
\newblock {\em The Journal of Chemical Physics}, 118(11):4775--4777, 02 2003. doi: {{%
10\hspace{.1pt}\discretionary{.}{%
}{.}\hspace{.4pt}1063\discretionary{/}{%
}{/}1\hspace{.1pt}\discretionary{.}{%
}{.}\hspace{.4pt}1558471}}


\bibitem{IDbMT}
D.~Morozov, K.~Beketayev, and G.~H. Weber.
\newblock Interleaving distance between merge trees.
\newblock In {\em Workshop on Topological Methods in Data Analysis and Visualization: Theory, Algorithms and Applications}, 2013.

\bibitem{TeiCMfPTaaID}
E.~Munch and A.~Stefanou.
\newblock The $\ell^\infty$-cophenetic metric for phylogenetic trees as an interleaving distance, 2018.

\bibitem{OtUoGHdfSC}
F.~Mémoli.
\newblock On the use of gromov--hausdorff distances for shape comparison.
\newblock In {\em Proceedings of the Eurographics Symposium on Point-Based Graphics}, pp. 81--90. The Eurographics Association, 2007.

\bibitem{pont2021wasserstein}
M.~Pont, J.~Vidal, J.~Delon, and J.~Tierny.
\newblock Wasserstein distances, geodesics and barycenters of merge trees.
\newblock {\em IEEE Transactions on Visualization and Computer Graphics}, 28(1):291--301, 2021.

\bibitem{CAotGHDaiAiNSM}
F.~Schmiedl.
\newblock Computational aspects of the gromov--hausdorff distance and its application in non-rigid shape matching.
\newblock {\em Discrete \& Computational Geometry}, 57(4):854--880, 2017.

\bibitem{sharma2025continuous}
M.~Sharma, T.~B. Masood, N.~H. List, I.~Hotz, and V.~Natarajan.
\newblock Continuous scatterplot and image moments for time-varying bivariate field analysis of electronic structure evolution.
\newblock {\em IEEE Transactions on Visualization and Computer Graphics}, 2025.

\bibitem{sridharamurthy2018edit}
R.~Sridharamurthy, T.~B. Masood, A.~Kamakshidasan, and V.~Natarajan.
\newblock Edit distance between merge trees.
\newblock {\em IEEE transactions on visualization and computer graphics}, 26(3):1518--1531, 2018.

\bibitem{SiSFT}
D.~M. Thomas and V.~Natarajan.
\newblock Symmetry in scalar field topology.
\newblock {\em IEEE Transactions on Visualization and Computer Graphics}, 17(12):2035--2044, 2011. doi: {{%
10\hspace{.1pt}\discretionary{.}{%
}{.}\hspace{.4pt}1109\discretionary{/}{%
}{/}TVCG\hspace{.1pt}\discretionary{.}{%
}{.}\hspace{.4pt}2011\hspace{.1pt}\discretionary{.}{%
}{.}\hspace{.4pt}236}}


\bibitem{MSDiSFbCC}
D.~M. Thomas and V.~Natarajan.
\newblock Multiscale symmetry detection in scalar fields by clustering contours.
\newblock {\em IEEE Transactions on Visualization and Computer Graphics}, 20(12):2427--2436, 2014. doi: {{%
10\hspace{.1pt}\discretionary{.}{%
}{.}\hspace{.4pt}1109\discretionary{/}{%
}{/}TVCG\hspace{.1pt}\discretionary{.}{%
}{.}\hspace{.4pt}2014\hspace{.1pt}\discretionary{.}{%
}{.}\hspace{.4pt}2346332}}


\bibitem{Thygesen2023}
S.~S. Thygesen, A.~I. Abrikosov, P.~Steneteg, T.~B. Masood, and I.~Hotz.
\newblock Level of detail visual analysis of structures in solid-state materials.
\newblock In {\em EuroVis Short Papers}, 2023.

\bibitem{GMMFaSA}
E.~F. Touli.
\newblock {\em Graphical Models: Mathematical Foundation and Statistical Analysis}.
\newblock Ph{D} dissertation, Department of Mathematics, Stockholm University, 2024.

\bibitem{FAfCGHaIDbT}
E.~F. Touli and Y.~Wang.
\newblock {FPT}-algorithms for computing {G}romov-{H}ausdorff and interleaving distances between trees.
\newblock {\em Journal of Computational Geometry}, 13(1):89--124, 2022.

\bibitem{wetzels2022branch}
F.~Wetzels, H.~Leitte, and C.~Garth.
\newblock Branch decomposition-independent edit distances for merge trees.
\newblock In {\em Computer Graphics Forum}, vol.~41, pp. 367--378. Wiley Online Library, 2022.

\bibitem{wetzels2025accelerating}
F.~Wetzels, H.~Leitte, and C.~Garth.
\newblock Accelerating computation of stable merge tree edit distances using parameterized heuristics.
\newblock {\em arXiv preprint arXiv:2501.05529}, 2025.

\bibitem{EEDEUMTM}
F.~Wetzels, T.~B. Masood, N.~H. List, I.~Hotz, and C.~Garth.
\newblock {Exploring Electron Density Evolution using Merge Tree Mappings}.
\newblock In C.~Tominski, M.~Waldner, and B.~Wang, eds., {\em EuroVis 2024 - Short Papers}. The Eurographics Association, 2024. doi: {{%
10\hspace{.1pt}\discretionary{.}{%
}{.}\hspace{.4pt}2312\discretionary{/}{%
}{/}evs\hspace{.1pt}\discretionary{.}{%
}{.}\hspace{.4pt}20241069}}


\bibitem{yan2022geometry}
L.~Yan, T.~B. Masood, F.~Rasheed, I.~Hotz, and B.~Wang.
\newblock Geometry-aware merge tree comparisons for time-varying data with interleaving distances.
\newblock {\em IEEE Transactions on Visualization and Computer Graphics}, 29(8):3489--3506, 2022.

\bibitem{yan2021scalar}
L.~Yan, T.~B. Masood, R.~Sridharamurthy, F.~Rasheed, V.~Natarajan, I.~Hotz, and B.~Wang.
\newblock Scalar field comparison with topological descriptors: Properties and applications for scientific visualization.
\newblock In {\em Computer Graphics Forum}, vol.~40, pp. 599--633. Wiley Online Library, 2021.

\bibitem{ASAoLMTfUV}
L.~Yan, Y.~Wang, E.~Munch, E.~Gasparovic, and B.~Wang.
\newblock A structural average of labeled merge trees for uncertainty visualization.
\newblock {\em IEEE Transactions on Visualization and Computer Graphics}, 26(1):832--842, Jan 2020. doi: {{%
10\hspace{.1pt}\discretionary{.}{%
}{.}\hspace{.4pt}1109\discretionary{/}{%
}{/}TVCG\hspace{.1pt}\discretionary{.}{%
}{.}\hspace{.4pt}2019\hspace{.1pt}\discretionary{.}{%
}{.}\hspace{.4pt}2934242}}


\bibitem{SMSRCULT}
K.~Zhang and T.~Jiang.
\newblock Some max snp-hard results concerning unordered labeled trees.
\newblock {\em Information Processing Letters}, 49(5):249--254, 1994. doi: {{%
10\hspace{.1pt}\discretionary{.}{%
}{.}\hspace{.4pt}1016\discretionary{/}{%
}{/}0020\discretionary{%
}{-}{-}0190\discretionary{%
}{(}{(}94\discretionary{)}{%
}{)}90062\discretionary{%
}{-}{-}0}}


\end{thebibliography}

\newpage

\appendix
\section{Pseudo-code for the First Method (Partial Agreement)}
\label{appen:appendixA}

\revision{Matrix $S$ in this paper is constructed using the following function:}
\begin{algorithmic}[1]
    \Statex 
    \Function{ConstructMatrixS}{$T, f$} 
        \State $n \gets$ number of unknown-labeled leaves in $T$
        \State $m \gets$ number of leaves in $T$
        \State Initialize matrix $S$ of size $n \times (m+1)$
        \For{each unknown-labeled leaf $v_i$ in $T$ (row $i$)}
            \For{each leaf $v_j$ in $T$ (column $j$)}
                \State $S(i,j) \gets f(\mathrm{LCA}(v_i, v_j)) - f(v_i)$
            \EndFor
            \State $S(i,n+1) \gets \sum_{j=1}^{n} S(i,j)$ \Comment{last column is row sum}
        \EndFor
        \State \Return $S$
    \EndFunction
\end{algorithmic}

\revision{In the case of partial agreement the function for constructing the distance matrix is as follows:}
\begin{algorithmic}[1]
  \Function{ConstructDistanceMatricesPA}{$T$} 
    \State $n \gets$ number of unknown-labeled leaves in $T$
    \State $m \gets$ number of known-labeled leaves in $T$
    \State Initialize matrix $D$ of size $n \times m$
    \For{each unknown-labeled leaf $v_i$ in $T$}
      \For{each leaf $v_j$ in $T$}
        \State $D(i, j) \gets$ shortest distance between $v_i$ and $v_j$
      \EndFor
    \EndFor
    \State \Return $D$
  \EndFunction
\end{algorithmic}

\revision{In the case of disagreement, the function for constructing the distance matrix for a subset $V$ of the vertices of tree $T$ is as follows:}

\begin{algorithmic}[1]
  \Function{ConstructDistanceMatricesDA}{$T$, $V$} 
    \State $n \gets$ number of $V$
    \State Initialize matrix $D$ of size $n \times n$
    \For{each unknown-labeled leaf $v_i$ in $V$}
      \For{each unknown-labeled leaf $v_j$ in $V$}
        \State $D(i, j) \gets$ shortest distance between $v_i$ and $v_j$ in $T$
      \EndFor
    \EndFor
    \State \Return $D$
  \EndFunction
\end{algorithmic}

\revision{The bipartite matching process for the two given distance matrices proceeds as follows:}

\begin{algorithmic}[1]
\Function{BipartiteMatching}{$D_1, D_2$}
    \State $n \gets$ the number of rows in $D_1$
    \State $m \gets$ the number of rows in $D_2$
    \State Initialize weight matrix $W$ of size $n \times m$
    \If{partial agreement case}
        \State Set edge weight: $W(i,j) \gets \| D_1(i,:) - D_2(j,:) \|_2$
    \ElsIf{disagreement case}
        \State Set edge weight: $W(i,j) \gets \left| \| D_1(i,:) \|_2 - \| D_2(j,:) \|_2 \right|$
    \EndIf
    \State $\mathit{matching}, V_\text{unmatched} \gets$ \Call{HungarianAlgorithm}{$G$}
    \State \Return $\mathit{matching}, V_{\text{unmatched}}$
\EndFunction
\end{algorithmic}

\revision{Below, you can find the pseudo-code for the induced matrix $M$ where $V_\text{match}$ is a subset of the vertices of the $T$ that is matched to vertices in another tree:}
\begin{algorithmic}[1]
\Function{InducedMatrix}{$T, f, V_{\text{match}}$}
    \For{each leaf $v_i$ in $V_{\text{match}}$}
        \For{each leaf $v_j$ in $V_{\text{match}}$}
            \State $M(i, j) \gets f(\mathrm{NCA}(v_i, v_j))$
        \EndFor
    \EndFor
    \State \Return $M$
\EndFunction
\end{algorithmic}

\revision{Below is the pseudo-code for the first algorithm (Efficient Leaf Matching Method) in the case of partial agreement; the disagreement case is analogous.}

\begin{algorithm}[H]
\caption{Efficient Leaf Matching  Method (Partial Agreement)}
\label{alg:heuristicfirstmethod}
\begin{algorithmic}[1]

\State \textbf{Input:} Two merge trees \( T_1^f \) and \( T_2^g \) with \( n_1 \) and \( n_2 \) unknown-labeled leaves, respectively.
\State \textbf{Output:} Estimated interleaving distance between \( T_1^f \) and \( T_2^g \).

\If{$n_1 > n_2$}
    \State $T_{\text{large}}^h \gets T_1^f$
\Else
    \State $T_{\text{large}}^h \gets T_2^g$
\EndIf

\State $S \gets$ \Call{ConstructMatrixS}{$T_{\text{large}}, h$}

\For{$i = 1$ \textbf{to} $|n_1 - n_2|$}
    \State $v_{j_{\min}} \gets \arg\min_j S(j, n)$ \Comment{Find the row index with minimum in last column}
    \State Append $v_{j_{\min}}$ to $V_{\text{trimmed}}$
\EndFor

\If{ \( n_1>0 \) and \( n_2>0 \) }
    \State $D_1 \gets$ \Call{ConstructDistanceMatricesPA}{$T_1$}
    \State $D_2 \gets$ \Call{ConstructDistanceMatricesPA}{$T_2$}
    \State $\text{matching}$, $V_\text{unmatched} \gets$ \Call{BipartiteMatching}{$D_1, D_2$} 
    \Comment{here $V_{\text{unmatched}} = \varnothing$}
    \State $V_{\text{match}} \gets$ leaves in the matching
    \State $M_1 \gets$ \Call{InducedMatrix}{$T_1, f, V_{\text{match}}$}
    \State $M_2 \gets$ \Call{InducedMatrix}{$T_2, g, V_{\text{match}}$}
\Else
    \State $V_{\text{known}} \gets$ known-labeled leaves 
    \State $M_1 \gets$ \Call{InducedMatrix}{$T_1, f, V_{\text{known}}$}
    \State $M_2 \gets$ \Call{InducedMatrix}{$T_2, g, V_{\text{known}}$}
\EndIf

\State $\ve = |M_1-M_2|_\infty$

\For{each leaf $v_i$ in $V_{\text{trimmed}}$}
    \State Remove leaf $v_i$ from the columns of $S$
    \State Remove the last column of $S$ (row sums)
    \State $\delta_i \gets \min_j S(i,j)$ \Comment{minimum value in the row corresponding to $v_i$}
\EndFor

\State \(\text{distance} \gets \max\left\{ \frac{1}{2} \max_i \delta_i, \, \varepsilon \right\}\)
\State \Return distance

\end{algorithmic}
\end{algorithm}

\section{Pseudo-code for the Second Method (Partial Agreement)}
\label{appen:appendixB}
\begin{algorithm}[H]
\caption{Modified Matching-Based Method}
\label{alg:heuristicsecondmethod}
\begin{algorithmic}[1]

\State \textbf{Input:} Two merge trees $T_1^f$ and $T_2^g$ with $n_1$ and $n_2$ unknown-labeled leaves
\State \textbf{Output:} Estimated interleaving distance between $T_1^f$ and $T_2^g$

\If{$n_1>0$ and $n_2>0$}
    \State $D_1 \gets$ \Call{ConstructDistanceMatricesPA}{$T_1$}
\State $D_2 \gets$ \Call{ConstructDistanceMatricesPA}{$T_2$}

 \State $\text{matching, $V_{\text{unmatched}}$} \gets$ \Call{BipartiteMatching}{$D_1, D_2$}
\State $V_{\text{match}} \gets$ leaves in the matching
    \State $M_1 \gets$ \Call{InducedMatrix}{$T_1, f, V_{\text{match}}$}
    \State $M_2 \gets$ \Call{InducedMatrix}{$T_2, g, V_{\text{match}}$}
\Else
    \State $V_{\text{known}} \gets$ known-labeled leaves 
    \State $M_1 \gets$ \Call{InducedMatrix}{$T_1, f, V_{\text{known}}$}
    \State $M_2 \gets$ \Call{InducedMatrix}{$T_2, g, V_{\text{known}}$}
\EndIf

\State $\ve = |M_1-M_2|_\infty$

\If{$n_1 > n_2$}
    \State $T_{\text{large}}^h \gets T_1^f$
\Else
    \State $T_{\text{large}}^h \gets T_2^g$
\EndIf

\State $n \gets |V_{\text{unmatched}}|$
\State $m \gets |T_{\text{large}} \setminus V_{\text{unmatched}}|$
\State Initialize matrix $D$ of size $n \times m$

\For{each unmatched leaf $v_i \in V_{\text{unmatched}}$}
    \For{$v_j$ in leaves($T_{\text{large}}\setminus V_{\text{unmatched}}$)}
        \State $S(i,j) \gets h(\mathrm{LCA}(v_i, v_j)) - h(v_i)$
    \EndFor
\EndFor

\For{each leaf $v_i$ in $V_{\text{unmatched}}$}
    \State $\delta_i \gets \min_j S(i,j)$
\EndFor

\State \(\text{distance} \gets \max\left\{ \frac{1}{2} \max_i \delta_i, \, \varepsilon \right\}\)
\State \Return distance

\end{algorithmic}
\end{algorithm}

\section{Additional Figures}
\begin{figure*}[!ht]
\centering
\includegraphics[width=17cm, trim=20 390 20 20, clip]{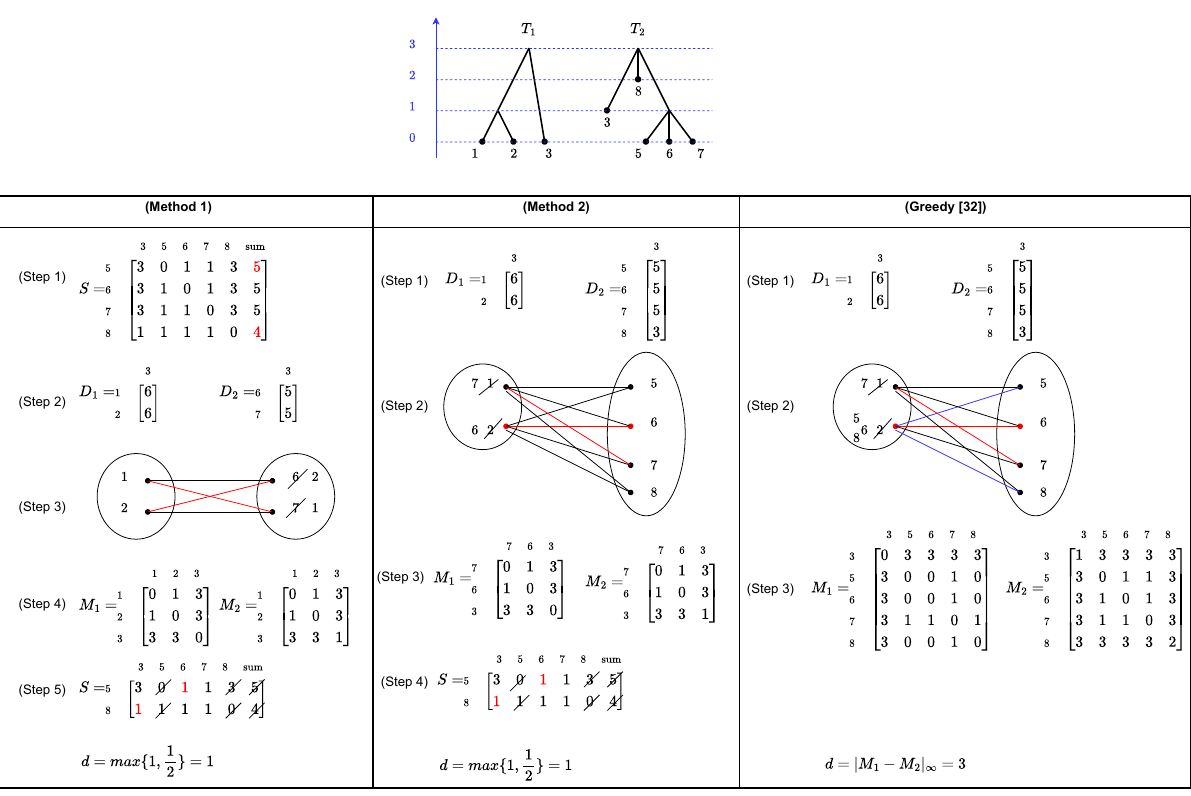}
\caption{
\revision{Comparison of the three methods using the same example. \emph{Method 1} (\textit{Efficient Leaf Matching Method}): In Step 1, the \( S \) matrix is constructed. Step 2 computes the distance matrices, followed by Step 3, where a bipartite graph is built. A minimum-weight maximum matching is applied, from which the induced matrix on the matched leaves is derived. Finally, the distance is estimated as $1$. \emph{Method 2} (\textit{Modified Matching-Based Method}): This method starts by computing the distance matrices, then constructs the bipartite graph. After solving the minimum-weight maximum matching problem, the induced matrix is formed. For any unmatched leaves, the corresponding \( \delta_i \) values are computed. The distance using this method is then estimated as $1$ as well. \emph{Method 3} (\textit{Greedy Method}, as discussed in \cite{ASAoLMTfUV}): This method also begins with computing the distance matrices, followed by bipartite graph construction. After solving the matching problem, unmatched leaves are assigned using a greedy strategy (illustrated with blue lines). Finally, the induced matrices for both trees are generated. In this example, using both the first and second methods yields the same distance of 
$1$, whereas the greedy method \cite{ASAoLMTfUV} produces a distance of 
$3$, highlighting the weakness of the greedy approach.}
}
\label{fig:exampleall}
\end{figure*}


\begin{figure*}[!ht]

    \raggedright
    \hspace{1.2em}
    \includegraphics[width=0.2\linewidth]{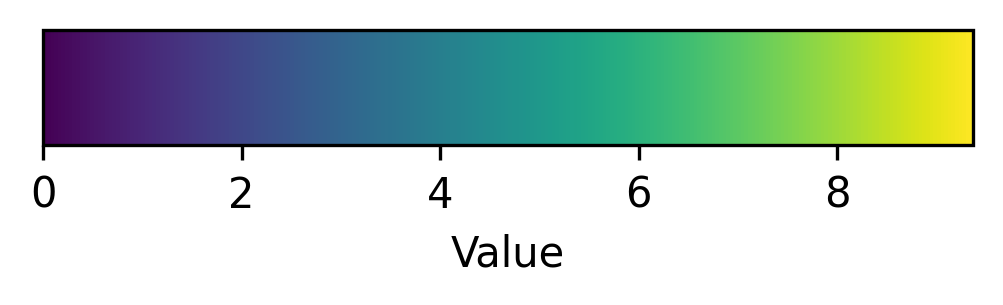} 
    \vspace{0.5em} 
    
    \centering

    \makebox[\linewidth][c]{%
        \hspace{-3em} 
        \begin{minipage}{0.8\linewidth}
            \centering
            \begin{tabular}{ccccc}
                \textbf{\MethodOne~~~~\qquad\qquad} & 
                \textbf{\MethodTwo ~~\qquad\qquad } & 
                \textbf{ Greedy~\cite{ASAoLMTfUV}\qquad\qquad} & 
                \textbf{ M1 vs Greedy\qquad\qquad} & 
                \textbf{M2 vs Greedy }
            \end{tabular}
        \end{minipage}
    }

    \vspace{0.5 em}

    \begin{tabular}{@{}c@{}c@{}}
    \raisebox{2cm}{\rotatebox[origin=c]{90}{\textbf{\texttt{Random\_50}}}} &
    \includegraphics[width=0.95\linewidth]{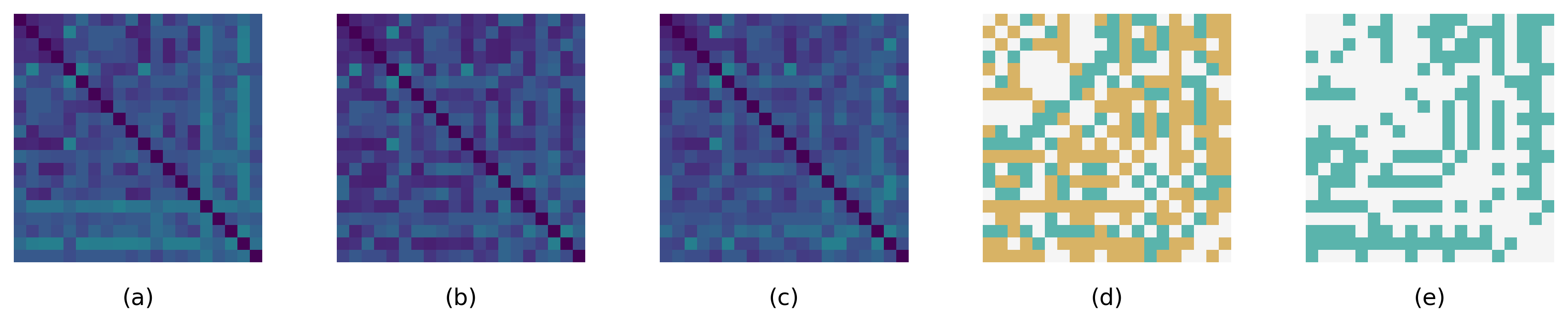} \\

    \raisebox{2cm}{\rotatebox[origin=c]{90}{\textbf{\texttt{Random\_100}}}} &
    \includegraphics[width=0.95\linewidth]{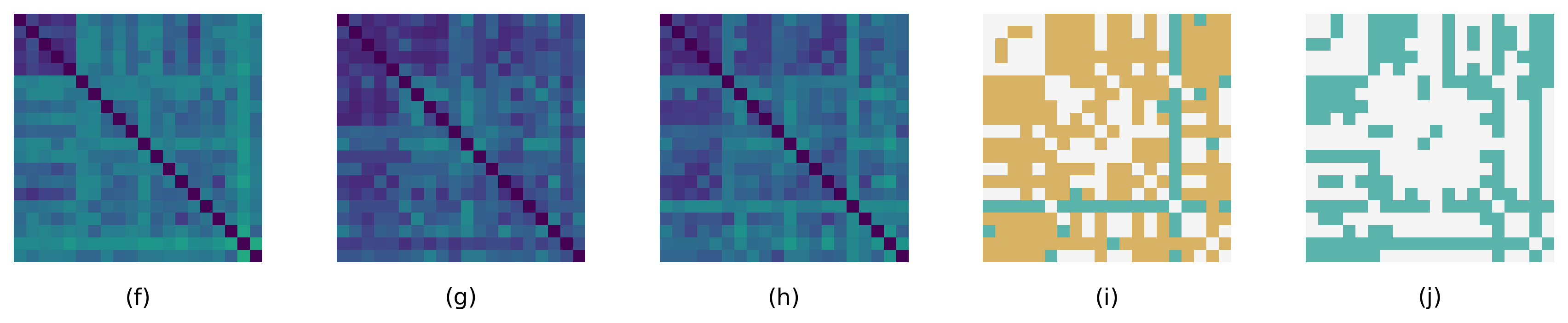} \\

    \raisebox{2cm}{\rotatebox[origin=c]{90}{\textbf{\texttt{Random\_200}}}} &
    \includegraphics[width=0.95\linewidth]{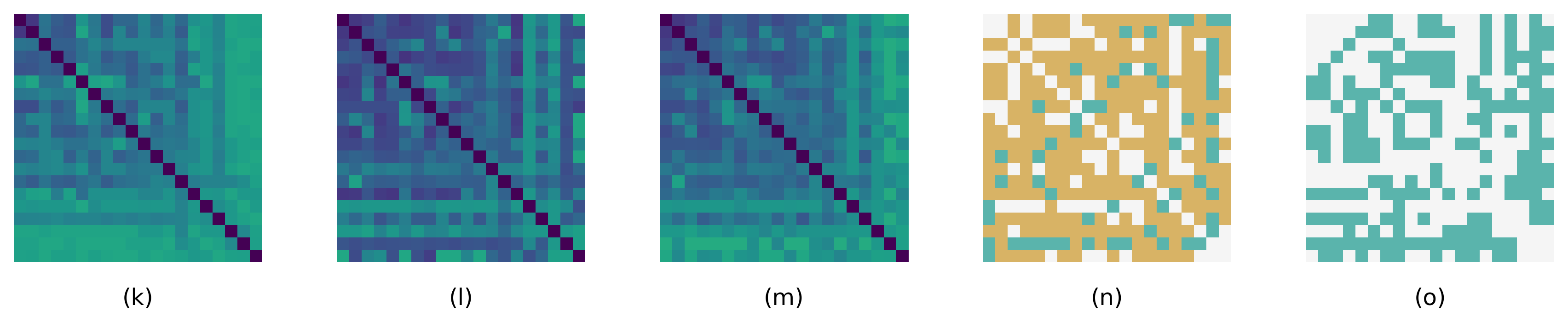} \\

    \raisebox{2cm}{\rotatebox[origin=c]{90}{\textbf{\texttt{Random\_500}}}} &
    \includegraphics[width=0.95\linewidth]{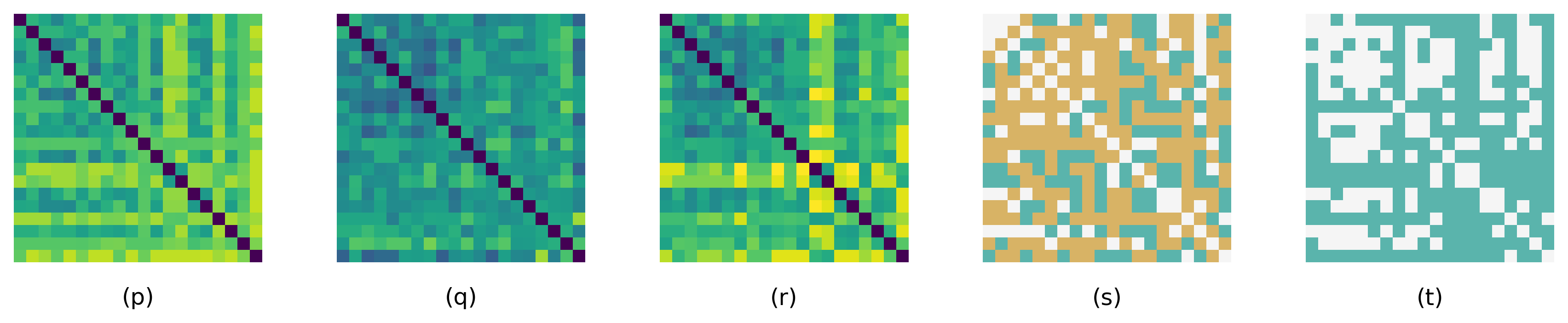} \\
    \end{tabular}

   \caption{
   \revision{Comparison of random trees using our first method, second method, and the greedy method described in \cite{ASAoLMTfUV}. The first row shows the distance matrices for trees with at most 50 vertices, the second row for trees with up to 100 vertices, the third row for trees with up to 200 vertices, and the final row for trees with up to 500 vertices. Similar to \cref{fig:real}, the fourth column shows regions of comparison between the first method and the greedy method \cite{ASAoLMTfUV}. Yellow areas indicate where the greedy method outperforms the first method, blue areas indicate where the first method outperforms the greedy method, and white areas indicate where both methods yield the same values. The last column shows the comparison between the second method and the greedy method: yellow areas mark regions where the greedy method performs better, blue areas mark regions where the second method performs better, and white areas mark regions where both methods yield the same values. In this case, no yellow areas are observed.}}
    \label{fig:random}
\end{figure*}

\begin{figure*}[!ht]
\centering
\includegraphics[width=15cm]{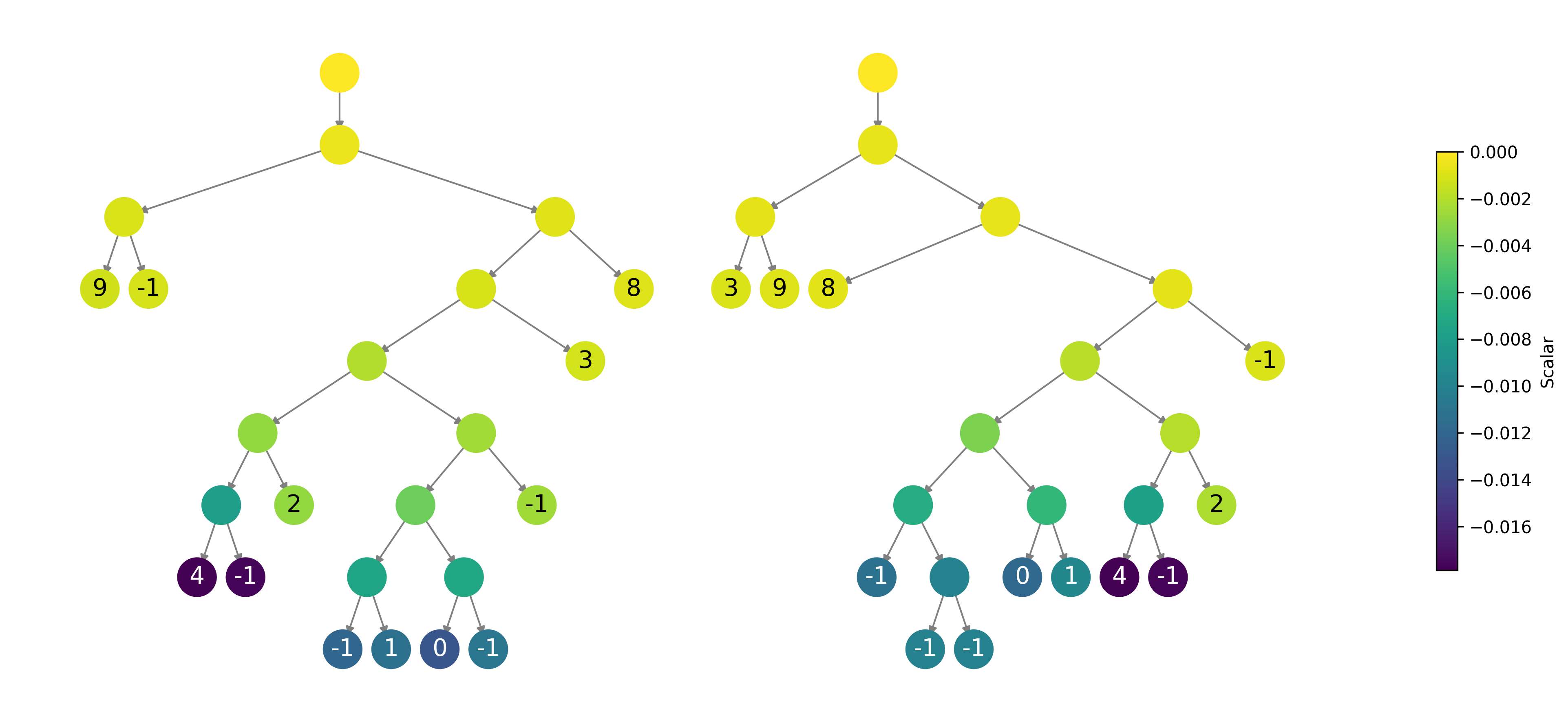}
\caption{
\revision{The tree on the left is obtained from \cref{fig:MVK}, while the tree on the right corresponds to the 10th time step of \StateTwoP. In this figure the leaves with the label $-1$ are unknown-labeled leaves.}}
\label{fig:trees}
\end{figure*}

\end{document}